# GENERATION AND MANIPULATION OF MULTI-COLOR STATIONARY LIGHT FIELD USING ELECTROMAGNETICALLY INDUCED TRANSPERANCY

S. A. Moiseev[1,2] and B. S. Ham[1]

[1]*The Graduate School of Information and Communications, Inha University, Incheon 402-751 S. Korea,*
[2]*Kazan Physical-Technical Institute of Russian Academy of Sciences, Kazan 420029 Russia*

E-mail: samoi@yandex.ru

Dynamic control of a weak quantum probe light pulse for the generation and quantum manipulations of a stationary multi-color (MC-) light field in a resonant coherent atomic medium using electromagnetically induced transparency is proposed. The manipulations have been analyzed based on the analytical solution of the adiabatic limit in the evolution of MC-light fields resulting from interaction of the slow probe light with the new fields generated in the nondegenerate multi-wave mixing scheme. We have found a critical stopping condition for the MC-light fields where the group velocity of light should reduce down to zero. Semiclassical dynamics and behavior of specific quantum correlations of the MC-light fields have been studied in detail for particular initial quantum states of the probe pulse. The stationary MC-field dynamics are treated in terms of dark MC-polariton states constructed for the studied multi-wave mixing processes. We have found the conditions for optimal manipulation of the MC-light while preserving the delicate quantum correlations of the initial probe light pulse. The quantum manipulations leading to the frequency and direction switching of the initial probe light pulse as well as to the quantum swapping of probe light into the new multi-frequency light fields have been proposed. The possibilities of the interaction time lengthening and enhancement of the electric field amplitudes of the stationary MC-light are also discussed for enhancement of the interactions with weak quantum fields in the spatially limited media.

## I. Introduction

The coherent interactions of the quantum light with resonant media has drawn much attention to the area of nonlinear quantum optics for the possibility of deterministic control of the light fields and for superior enhancement of nonlinear interactions [1]. Such issues are most important for quantum information processes, which especially need repeated non-destructive measurements of physical observables and entanglement control of the light fields [2]. The most successful experimental results in this area were achieved using a cavity-QED technique which exploits the cavity-lengthening interaction time [3]. It would be important to realize these results for a traveling light schemes, which promise important possibilities for control of the light-matter interactions. Recently electromagnetically induced transparency (EIT) [4] has been successfully applied for the enhancement of light-media interaction and for a considerable lengthening of the interaction time [5, 6]. EIT is a quantum optical phenomenon of the resonant interaction between the light and coherent medium, where the additional control laser field modifies the absorption and refractive index of the medium resulting in transparency of the medium even to a resonant probe light. The modified refractive index of the medium results in the ultra-slow group velocity of the probe light pulse observed in different media [7-13]. Quantum control of the interactions with slow light fields can be used for quantum switching [14], quantum memory [15, 16], deterministic entanglement [17], and for quantum computing [18]. Highly efficient nondegenerate four-wave mixing generations have also been shown using EIT-induced giant Kerr nonlinearity [19], which is likely to provide broad potential for using a very weak optical field to obtain a π-phase shift [20-22], where a quantum superposition such as Schrödinger's cat can be realized [20,22].

The quantum dynamics of the slow light field in the EIT condition demonstrates unusual and often unpredictable behavior so the slow light phenomenon still attracts a deep physical interest [23]. Very recently, a new effect - stationary light using EIT-induced slow light has been observed in the hot gas with Λ-type atomic medium using a standing-wave grating formation caused by the counter-propagating intensive laser fields [24]. The stationary light is different from the quantum memory phenomenon [15,16], where a complete quantum conversion process between the light and media is not involved therefore the stationary light field can interact for a longer duration with individual localized atoms or another light field in the spatially limited media [25]. Thus, one can principally apply the stationary light for temporally lengthening manipulation of probe weak quantum light fields that is particularly similar to a cavity-QED, which provides deterministic quantum information processes especially for the quantum nondemolition measurement where low nonlinearity is a critical problem. Nevertheless it should be noted here that such nonlinear interactions with single photon fields should satisfy strong demands to a spatial compression of the light beams and spectral parameters of the interacting light fields and atomic medium so the problem still needs further research to provide a stronger coupling constant of photon-photon interactions in the coherent media [26].

Recently the quantum control of the stationary slow light has been proposed for the generation and control of the two- and three-color stationary light [27] with the possibility of an effective wavelength conversion of initial quantum probe light. Transition to the quantum control of stationary multi-frequency fields opens a new spectral freedom degree for quantum manipulation of the light fields which is important issue for multi-frequency interface and quantum communication. In this paper we present a generalized theoretical approach to the stationary light control based on the nondegenerate multi-wave mixing processes in a resonant multi-Λ scheme (see Fig. 1). In the framework of the adiabatic approach to multi-waves interaction in the coherent medium, we analytically study the generation of new light fields with arbitrary frequency components from the initial probe light field $E_1$ launched into the medium

(below we call the generated coupled light fields by a *multi-color* (MC-) light fields). We analyze the quantum manipulations of MC-field. In particular we have found that the generated light fields are coupled undividedly with each other in the MC-field by united group velocity and similar spatial envelopes which can be manipulated and even completely stopped by adjusting the laser field amplitudes.

The proposed quantum control of the multi-wave dynamics opens the possibilities for: 1) controllable lengthening of the stationary MC light evolution in the medium, 2) enhancement of the electric field amplitudes for the chosen MC-light field components and 3) controllable frequency tuning of the different spectral field components and transformation of the initial quantum correlations of probe light to the new one or many traveling coupled light fields. We have found the optimal spectral conditions where such quantum manipulations of the MC-light fields take place with minimum distortions of the original parameters of probe pulse energy and quantum correlations. For this purpose we have analyzed in detail the semiclassical dynamics of MC-light evolution for the coherent state of the probe light and have examined the behavior of specific quantum correlations for spectrally correlated two photon (EPR) probe field. In the conclusion, we summarize the obtained results and possible applications of the proposed manipulations with the weak probe quantum light pulse.

## II. Adiabatic Theory of MC light

Figure 1 shows an energy level diagram of the interaction between all the quantum light fields $M_+ + M_-$ in the coherent resonance medium derived by the $M_+ + M_-$ control intensive laser fields. We assume the quantum fields $\hat{E}_m$ ($m \in [1,...,M_+]$) and the control fields $\Omega_m$ propagate along z-direction whereas the other quantum fields $\hat{E}_n$ ($n \in [1,...M_-]$) and their control fields $\Omega_n$ have an opposite backward –z direction. Such directions of the field waves correspond to the effective generation and strong interaction between the waves at the phase-matching condition. As seen in Fig. 1, the control fields $\Omega_m$ and $\Omega_n$ have the frequencies close to the frequencies of the atomic transitions between the state |2⟩ and the excited states |m⟩ and |n⟩ (where $\Omega_m(t) = \Omega_{m,o}(t)e^{i\phi_m}$ and $\Omega_n(t) = \Omega_{n,o}(t)e^{i\phi_n}$ are the Rabi frequencies on the atomic transitions $|2\rangle \leftrightarrow |m\rangle$ and $|2\rangle \leftrightarrow |n\rangle$, $\phi_{m(n)}$ are the constant phases of the fields).

We assume that initially all the atoms stay on the ground level 1 and only the one probe light field $\hat{E}_l$ enters into the medium at t=0 and the atoms are driven by one intensive control laser field $\Omega_l(t \approx 0) \neq 0$. The initial state of the light and medium in the interaction picture is given by $|\varphi(t<<0)\rangle_{in} = |\varphi_l\rangle|1\rangle_{atoms}$, where $|1\rangle_{atoms} = \prod_j |1_j\rangle$ and $|\varphi_l\rangle$ are a ground state of the atoms and initial state of the probe weak light field $\hat{E}_l$. The two fields $\hat{E}_l$ and $\Omega_l$ create the atomic coherent wave $\langle \hat{P}_{12} \rangle$ between level 1 and 2 which follows the slowly propagating light pulse with the wave vector $k_l - K_l$ (where we assume $k_{m,n}$ and $K_{m,n}$ are wave vectors of the fields $\hat{E}_{m,n}$ and $\Omega_{m,n}$). After the complete entrance of the probe pulse into the medium we propose to switch on the additional control laser fields for t>$t_o$ so $\Omega_m(t>t_o) \neq 0$ and $\Omega_n(t>t_o) \neq 0$. Such switching processes can be easily realized due to the very slow group velocity of the probe light. The additional control fields $\Omega_{m,n}$ will create the new atomic polarization waves (with wave vectors $k_{m,n} = k_l - K_l + K_{m,n}$) each of them generate the new light field $\hat{E}_{m,n}$ intensively at the phase matching condition. Thus the present multi-color light control utilizes a single color traveling probe light to create $M_+ + M_-$ spectral components of stationary (multi-color) light using many double $\Lambda$ EIT scheme, in which all the spectral components become absorption free. All the quantum fields $\hat{E}_l$, $\hat{E}_{m(\neq l)}$ and $\hat{E}_n$ will interact intensively with each others in the optically dense medium via the huge joint atomic coherence wave $\langle \hat{P}_{12} \rangle$ whereas the optical atomic coherences $\langle \hat{P}_{m(n),2} \rangle$ corresponding to the individual light fields $\hat{E}_{m,n}$ will follow adiabatically the dynamics of coherence wave $\langle \hat{P}_{12} \rangle$. Therefore all fields $\hat{E}_{m,n}$ become indivisibly coupled with each other, which changes its properties dramatically in comparison to behavior of the initial traveling probe light field. The optimum conditions of the perfect quantum evolution of MC-light are studied below.

For a theoretical analysis of the evolution of MC-light fields we use a one dimensional approach for the quantum fields of light beams $\hat{E}_m(t,z) = \sqrt{\hbar\omega_m/(2\varepsilon_o S)}\hat{A}_m(t,z)e^{-i(\omega_m t - k_m z)}$ and $\hat{E}_n(t,z) = \sqrt{\hbar\omega_n/(2\varepsilon_o S)}\hat{A}_n(t,z)e^{-i(\omega_n t + k_n z)}$, where $\hat{A}_{m;(n)}$ are slowly varying operators corresponding to the forward waves (index m) which are resonant with atomic transitions $|1\rangle \rightarrow |m\rangle$ and the wave index n stands for the backward fields resonant to the transitions $|1\rangle \rightarrow |n\rangle$, respectively (see Fig.1) (where $[\hat{A}_p(t,z), \hat{A}_q(t,z')] = \delta_{p,q}\delta(z-z')$, $p,q$ are the common indexes corresponding to all the fields including their polarization, $\hbar$ is the Planck's constant, $\varepsilon_o$ is the electric permittivity, S is a cross section of light beams [1, 28]). In the interaction picture, the Hamiltonian of the quantum fields with the atoms which are driven by the additional control laser fields $\Omega_m(t)$ and $\Omega_n(t)$ is

$$\hat{H}(t) = \hat{V}_f(t) + \hat{V}_b(t) + H.C., \tag{1a}$$

$$\hat{V}_f(t) = -\hbar \sum_j \{\sum_{m=1}^{M_+} g_m \hat{A}_m(t,z_j) e^{-i(\Delta_m t - k_m z_j)} \hat{P}_{m1}^j + \sum_{m=1}^{M_+} \Omega_m(t) e^{-i(\Delta_m t - (k_m - k_o)z_j)} \hat{P}_{m2}^j\}, \tag{1b}$$

$$\hat{V}_b(t) = -\hbar \sum_j \{\sum_{n=1}^{M_-} g_n \hat{A}_n(t,z_j) e^{-i(\Delta_n t + k_n z_j)} \hat{P}_{n1}^j + \sum_{n=1}^{M_-} \Omega_n(t) e^{-i(\Delta_n t + (k_n - k_o)z_j)} \hat{P}_{n2}^j\}, \tag{1c}$$

where $\hat{P}_{ll'}^j = |l\rangle_{jj}\langle l'|$ are an operator of the j-th atom, $g_l = \wp_{l,1}\sqrt{\omega_{l,1}/(2\varepsilon_o \hbar S)}$ is a coupling constant of photons with atoms, $\wp_{l',l}$ is a dipole moment for the transition between the states $|l'\rangle \to |l\rangle$, $v_m = \omega_{m1} + \Delta_m$, and $v_n = \omega_{n1} + \Delta_n$, $k_o = \omega_{21}/c$.

We study the quantum dynamics in the adiabatic limit and show that the quantum state of the weak quantum fields $\hat{E}_{m,n}$ and atoms follows near to the dark states of the Hamiltonian in Eq. (1). In the spirit of the single-color slow light phenomenon [5, 6, 29], we are able to construct a more general family of the dark- and bright MC-polariton states corresponding to the Hamiltonian in Eq. (1). In particular the arbitrary dark MC-polariton states with $n$ excited polaritons are found from the following condition $H(t)|D_n(t)\rangle = 0$:

$$|D_n(t)\rangle_{f+a} = \int \frac{dz_1,...,dz_n}{\sqrt{n!}} \phi_{f+a}(t,z_1,...,z_n) \hat{\Psi}_{f+a}^+(t,z_1),...,\hat{\Psi}_{f+a}^+(t,z_n)|0\rangle, \tag{2a}$$

$$\hat{\Psi}_{f+a}^+(t,z) = \langle D_{f+a}\rangle^{-1/2} \{\sum_{m=1}^{M_+}(\Omega_m(t)/g_m)\hat{A}_m^+(t,z) + \sum_{n=1}^{M_-}(\Omega_n(t)/g_n)\hat{A}_n^+(t,z) - \sqrt{n_o S}\hat{P}_{21}(t,z)\}, \tag{2b}$$

where $\langle D_{f+a}\rangle = n_o S + \sum_{m=1}^{M_+}\left|\frac{\Omega_m(t,z)}{g_m}\right|^2 + \sum_{n=1}^{M_-}\left|\frac{\Omega_n(t,z)}{g_n}\right|^2$, $\int dz_1,...,dz_n |\phi_{f+a}(t,z_1,...,z_n)|^2 = 1$,

$\hat{P}_{pq}(t,z) = (n_o S)^{-1/2}\sum_j \hat{P}_{pq}^j(t)\delta(z-z_j)$, $n_o$ is an atomic density and we have assumed $k_o L \ll 1$, L is a longitudinal size of the medium.

General dark MC-polariton state will be:

$$|D(t)\rangle_{f+a} = \sum_{n=0} C_n(t)|D_n(t)\rangle_{f+a}, \tag{2c}$$

where $\sum_n |C_n(t)|^2 = 1$. We note the field operators $\hat{\Psi}_{f+b}^+(t,z)$ satisfy the following boson commutation relations

$$[\hat{\Psi}_{f+a}^+(t,z), \hat{\Psi}_{f+a}^+(t,z')] = 0, \tag{3a}$$

$$[\hat{\Psi}_{f+a}(t,z), \hat{\Psi}^+_{f+a}(t,z')] = \delta(z-z'),\tag{3b}$$

where we have used the property $[\hat{P}_{pq}(t,z), \hat{P}_{q'p'}(t,z')] = \delta_{p1}\delta_{qq'}\delta(z-z')$ which takes place in the optically dense medium in the limit of weak quantum fields $\hat{E}_{m,n}$ when most of the atoms stay on the ground level 1 similar to [29].

We note that from Eq. (2b) we have for $t < t_o$ a well-known operator

$$\hat{\Psi}^+_{f+a}(t<t_o,z) = \cos\vartheta_l(t)\hat{A}^+_l(t,z) - \sin\vartheta_l(t)\hat{P}_{21}(t,z),\tag{4}$$

corresponding to single color slow light polariton field [29] (where $\tan\vartheta_l(t) = \sqrt{n_o S}g_l/\Omega_l(t)$). The bright MC-polariton states are also easily constructed for the classical control laser fields $\Omega_{m,n}(t)$ so the arbitrary state of the quantum light fields and atoms can be described through the arbitrary quantum superposition of all the dark and bright MC-polariton states, that will be studied in detail elsewhere. Below we study the coherent dynamics directly solving the Heisenberg equations of Hamiltonian in Eq. (1) and comparing the temporal and spatial properties of the light-medium evolution with its properties in the dark MC- states of Eq. (2). Since the MC-dark states in Eq. (2c) are decoupled from excited atomic states |m⟩ and |n⟩ for the Hamiltonian of Eq. (1) therefore the states play an important role in nonabsorbing evolution of all the interacting fields $\hat{E}_{m,n}$. As seen from Eq. (2b) the united operator $\hat{\Psi}^+_{f+a}(t,z)$ couples all the light fields $\hat{A}^+_{m(n)}(t,z)$ and atomic coherence $\hat{P}_{21}(t,z)$ in one common quantum MC-polariton field. Using Eq. (2b) we find the following commutation relations of the light fields with the polariton field

$$[\hat{A}_{m(n)}(t,z'), \hat{\Psi}^+_{f+a}(t,z)] = \Omega_{m(n)}(t)(g_{m(n)}\langle D_{f+a}\rangle^{1/2})^{-1}\delta(z-z').\tag{5}$$

We obtain the following relations for the quantum average of the field operators $\hat{\Psi}_m(t,z) = e^{ik_o z}\sqrt{n_o S}g_m\hat{A}_m(t,z)/\Omega_m(t)$ and $\hat{\Psi}_n(t,z) = e^{-ik_o z}\sqrt{n_o S}g_n\hat{A}_n(t,z)/\Omega_n(t)$ taking into account the optically dense medium $n_o S \gg \sum_{m=1}^{M_+}\left|\frac{\Omega_m(t)}{g_m}\right|^2 + \sum_{n=1}^{M_-}\left|\frac{\Omega_n(t)}{g_n}\right|^2$ in the dark MC- states (2c)

$$e^{-ik_o z}\langle\hat{\Psi}_m(t,z)\rangle = e^{ik_o z}\langle\hat{\Psi}_n(t,z)\rangle = \langle\hat{\Psi}_{f+a}(t,z)\rangle.\tag{6}$$

Such simple relations can be used for analysis of nearness of the studied quantum fields the pure unitary evolution of the dark MC-states. Nearness of the quantum evolutions of the fields $\hat{\Psi}_{m(n)}(t,z)$ in the adiabatic limit also makes the analysis of quantum dynamics convenient in terms of these operators.

Using the Hamiltonian (1) we initially derive the Heisenberg equations for the atomic operators $\hat{S}_{1m}^j = \hat{P}_{1m}^j e^{i(\Delta_m t - k_m z_j)}$, $\hat{S}_{1n}^j = \hat{P}_{1n}^j e^{i(\Delta_n t + k_n z_j)}$ and $\hat{P}_{12}^j$ by adding the relaxation constants of the atomic coherences $\gamma_{1(2),m}$, $\gamma_{1(2),n}$ and level populations $w_{1p}$ and $w_{2p}$, as well the Langevin forces $\hat{F}_{pq;(0)}^j(t)$ associated with the relaxation [1, 30]:

$$\tfrac{\partial}{\partial t}\hat{S}_{1m}^j = -\tilde{\gamma}_{1m}\hat{S}_{1m}^j + ig_m \hat{A}_m(t,z_j)(\hat{P}_{11}^j - \hat{P}_{mm}^j) + i\Omega_m e^{-ik_o z_j}\hat{P}_{12}^j + \hat{F}_{1m;(1)}^j(t), \tag{7a}$$

$$\tfrac{\partial}{\partial t}\hat{S}_{m2}^j = -\tilde{\gamma}_{2m}^*\hat{S}_{m2}^j - ig_m^* \hat{A}_m^+(t,z_j)\hat{P}_{12}^j + i\Omega_m^* e^{ik_o z_j}(\hat{P}_{mm}^j - \hat{P}_{22}^j) + \hat{F}_{m2;(1)}^j(t), \tag{7b}$$

$$\tfrac{\partial}{\partial t}\hat{S}_{1n}^j = -\tilde{\gamma}_{1n}\hat{S}_{1n}^j + ig_n \hat{A}_n(t,z_j)(\hat{P}_{11}^j - \hat{P}_{nn}^j) + i\Omega_n e^{ik_o z_j}\hat{P}_{12}^j + \hat{F}_{1n;(1)}^j(t), \tag{7c}$$

$$\tfrac{\partial}{\partial t}\hat{S}_{n2}^j = -\tilde{\gamma}_{2m}^*\hat{S}_{m2}^j - ig_m^* \hat{A}_m^+(t,z_j)\hat{P}_{12}^j + i\Omega_m^* e^{-ik_o z_j}(\hat{P}_{nn}^j - \hat{P}_{22}^j) + \hat{F}_{n2;(1)}^j(t), \tag{7d}$$

$$\tfrac{\partial}{\partial t}\hat{P}_{12}^j = -\gamma_{12}\hat{P}_{12}^j + i\sum_{m=1}^{M_+}\Omega_m^*(t)e^{ik_o z_j}\hat{S}_{1m}^j + i\sum_{n=1}^{M}\Omega_n^*(t)e^{-ik_o z_j}\hat{S}_{1n}^j - ig_m \hat{A}_m(t,z_j)\hat{S}_{m2}^j - ig_n \hat{A}_n(t,z_j)\hat{S}_{n2}^j + \hat{F}_{12;(0)}^j(t). \tag{7e}$$

$$\tfrac{\partial}{\partial t}\hat{P}_{22}^j = -\gamma_{22}\hat{P}_{nn}^j + i(\Omega_m^* e^{ik_o z_j}\hat{S}_{2m}^j + \Omega_n^* e^{-ik_o z_j}\hat{S}_{2n}^j) + \sum_p w_{2p}\hat{P}_{pp}^j + \hat{F}_{22}^{(1)}(t,z_j) - w_{12}(\hat{P}_{22}^j - \hat{P}_{11}^j), \tag{7f}$$

where $\tilde{\gamma}_{m,n} = \gamma_{m,n} - i\Delta_{m,n}$, $\hat{F}_{1m;(1)}^j(t) = e^{i[\Delta_m t - k_m z_j]}\hat{F}_{1m;(0)}^j(t)$, $\hat{F}_{m2;(1)}^j(t) = e^{-i[\Delta_m t - k_m z_j]}\hat{F}_{m2;(0)}^j(t)$, $\hat{F}_{1n;(1)}^j(t) = e^{i[\Delta_n t + k_n z_j]}\hat{F}_{1n;(0)}^j(t)$; $\hat{F}_{n2;(1)}^j(t) = e^{-i[\Delta_n t + k_n z_j]}\hat{F}_{n2;(0)}^j(t)$ and the fluctuations forces satisfy usual properties [1, 30]:

$$\langle \hat{F}_{pq;(0)}^j(t)\rangle = 0, \tag{8a}$$

$$\langle \hat{F}_{1m(n);(0)}^j(t)\hat{F}_{m(n)2;(0)}^j(t')\rangle = (\gamma_{1,m(n)} + \gamma_{m(n),2} - \gamma_{12})\langle \hat{P}_{12}^j(t)\rangle \delta(t-t'), \tag{8b}$$

$$\langle \hat{F}_{1m(n);(0)}^j(t)\hat{F}_{m(n)1;(0)}^j(t')\rangle = \{2\gamma_{1m(n)}\langle \hat{P}_{11}^j(t)\rangle + \sum_p w_{1p}\langle \hat{P}_{pp}^j(t)\rangle + w_{12}(\langle \hat{P}_{22}^j(t)\rangle - \langle \hat{P}_{11}^j(t)\rangle)\}\delta(t-t'), \tag{8c}$$

$$\langle \hat{F}_{2m(n);(0)}^j(t)\hat{F}_{m(n)2;(0)}^j(t')\rangle = \{2\gamma_{2m(n)}\langle \hat{P}_{22}^j(t)\rangle + \sum_p w_{2p}\langle \hat{P}_{pp}^j(t)\rangle - w_{12}(\langle \hat{P}_{22}^j(t)\rangle - \langle \hat{P}_{11}^j(t)\rangle)\}\delta(t-t'). \tag{8d}$$

Due to the weakness of the quantum fields $\hat{A}_{m;(n)}(t,z)$ in Eqs (7) we ignore the operators $\hat{P}_{qq}^j$ ($q \neq 1$) assuming negligible weak populations of excited states $|m\rangle$ and $|n\rangle$ so $\langle \hat{P}_{mm}^j\rangle \cong \langle \hat{P}_{nn}^j\rangle \cong 0$ [1]. Using Hamiltonian in Eq. (1) we also add Eqs. (7a)-(7f) by the Heisenberg equations for the field operators $\hat{A}_{m,n}$:

$$(\tfrac{\partial}{c\partial t} + \tfrac{\partial}{\partial z})\hat{A}_m(t,z) = i(\sqrt{n_o S}g_m^*/c)\hat{S}_{1m}(t,z), \tag{9a}$$

$$(\tfrac{\partial}{c\partial t} - \tfrac{\partial}{\partial z})\hat{A}_n(t,z) = i(\sqrt{n_o S}g_n^*/c)\hat{S}_{1n}(t,z), \tag{9b}$$

where $\hat{S}_{1m(n)}(t,z) = (n_o S)^{-1/2}\sum_j \hat{S}_{1,m(n)}^j(t)\delta(z-z_j)$.

In the first step of the solution we assume a large enough temporal duration $\delta t_o$ of the initial probe light pulse $\tilde{\gamma}_{m,n}\delta t_o >> 1$ and find the following approximate solutions of Eqs (7b) and

(7d) $\hat{S}_{m2}^j \cong \tilde{\gamma}_{m2}^{-1}\hat{F}_{m2;(2)}^j(t)$, $\hat{S}_{n2}^j \cong \tilde{\gamma}_{n2}^{-1}\hat{F}_{n2;(2)}^j(t)$ (where $\hat{F}_{pq;(2)}^j(t) = \tilde{\gamma}_{pq}\int_{t_o}^t dt'\exp\{-\tilde{\gamma}_{pq}(t-t')\}\hat{F}_{pq;(1)}^j(t')$ ) and

similarly we find in Eqs. (7a) and (7c):

$$\hat{S}_{1m}^j(t) \cong i\tilde{\gamma}_{1m}^{-1}\{g_m\hat{A}_m(t,z_j) + \Omega_m e^{-ik_o z_j}\hat{P}_{12}^j(t)\} + \tilde{\gamma}_{1m}^{-1}\hat{F}_{1m;(2)}^j(t), \tag{10a}$$

$$\hat{S}_{1n}^j(t) \cong i\tilde{\gamma}_{1n}^{-1}\{g_n\hat{A}_n(t,z_j) + \Omega_n e^{ik_o z_j}\hat{P}_{12}^j(t)\} + \tilde{\gamma}_{1n}^{-1}\hat{F}_{1n;(2)}^j(t), \tag{10b}$$

Putting $\hat{S}_{m2}^j(t)$, $\hat{S}_{n2}^j(t)$, $\hat{S}_{1m}^j(t)$ and $\hat{S}_{1n}^j(t)$ to Eq. (7e) we get

$$\tfrac{\partial}{\partial t}\hat{P}_{12}^j = -\mu\hat{P}_{12}^j - N^{-1/2}G(\hat{\Psi}(t,z_j)) + \hat{F}_{12;(\Sigma 1)}^j(t), \tag{11a}$$

$$\mu(t) = \{\gamma_{12} + \sum_{m=1}^{M_+}\Gamma_m(t) + \sum_{n=1}^{M_-}\Gamma_n(t)\}, \tag{11b}$$

$$G(\hat{\Psi}(t,z)) = \{\sum_{m=1}^{M_+}\Gamma_m(t)\hat{\Psi}_m(t,z) + \sum_{n=1}^{M_-}\Gamma_n(t)\hat{\Psi}_n(t,z)\}, \tag{11c}$$

$$\hat{F}_{12;(\Sigma 1)}^j(t) = i\{\sum_{m=1}^{M_+}[e^{ik_o z_j}\tfrac{\Omega_m^*(t)}{\tilde{\gamma}_{1m}}\hat{F}_{1m;(2)}^j(t) - \tfrac{g_m\hat{A}_m(t,z_j)}{\tilde{\gamma}_{m2}}\hat{F}_{m2;(2)}^j(t)] + \sum_{n=1}^{M_-}[e^{-ik_o z_j}\tfrac{\Omega_n^*(t)}{\tilde{\gamma}_{1n}}\hat{F}_{1n;(2)}^j(t) - \tfrac{g_n\hat{A}_n(t,z_j)}{\tilde{\gamma}_{n2}}\hat{F}_{n2;(2)}^j(t)]\} + \hat{F}_{12;(0)}^j(t), \tag{11d}$$

where $\Gamma_{m(n)}(t) = \Omega_{m(n);o}^2(t)/\tilde{\gamma}_{m(n)}$. We note that the fluctuation forces $\hat{F}_{m2;(2)}^j(t)$ and $\hat{F}_{n2;(2)}^j(t)$ in Eq. (11d) can be ignored in comparison to the forces $\hat{F}_{1m;(2)}^j(t)$ and $\hat{F}_{1n;(2)}^j(t)$ because of the weak quantum fields $g_{m(n)}\langle\hat{A}_{m(n)}(t,z)\rangle << \Omega_{m(n)}$ [16, 32, 33].

In the next step we write the formal solution of Eq. (11a) which correspond to the adiabatic limit of the atomic evolution

$$\hat{P}_{12}^j = \hat{P}_{12;(0)}^j + \hat{P}_{12;(1)}^j + \mu^{-1}\hat{F}_{12,(\Sigma 2)}^j, \tag{12a}$$

$$\hat{F}_{12,(\Sigma 2)}^j(t) = \mu(t)\int_{t_o}^t dt'\exp\{-\int_{t'}^t dt''\mu(t'')\}\hat{F}_{12,(\Sigma 1)}^j(t'), \tag{12b}$$

where $\hat{P}_{12;(0)}^j \cong -N^{-1/2}G(\hat{\Psi}(t,z_j))/\mu(t)$, $\hat{P}_{12;(1)}^j(t) \cong N^{-1/2}(\partial/\mu\partial t)[G(\hat{\Psi}(t,z_j)/\mu(t)]$ and $\hat{P}_{12;(1)}^j << \hat{P}_{12;(0)}^j$. Putting Eqs. (12a) and (12b) in Eqs. (9) and (10) and accepting the conditions of the slow-light propagation for $t<t_o$ with initial group velocity $v_l$ of the probe field $\hat{A}_l$ (we introduce the notations $v_{m,n}(t) = c\Omega_{m,n}^2(t)/Ng_{m,n}^2 << c$) we finally obtain the following system of equations for the field operators $\hat{\Psi}_m(t,z)$ and $\hat{\Psi}_n(t,z)$:

$$(\tfrac{\partial}{\partial z} - ik_o)\hat{\Psi}_m = -\xi_m\{\hat{\Psi}_m - (1-\tfrac{\partial}{\mu\partial t})[\tfrac{1}{\mu}G(\hat{\Psi})]\} - (\xi_m N/\mu)\hat{F}_{1m}^{(3)}, \tag{13a}$$

$$(\tfrac{\partial}{\partial z} + ik_o)\hat{\Psi}_n = \xi_n\{\hat{\Psi}_n - (1-\tfrac{\partial}{\mu\partial t})[\tfrac{1}{\mu}G(\hat{\Psi})]\} + (\xi_n N/\mu)\hat{F}_{1n}^{(3)}, \tag{13b}$$

where the fluctuation forces are

$$\hat{F}_{1m}^{(3)}(t,z) = N^{-1/2}\sum_{j}\{\hat{F}_{12,(\Sigma 2)}^{j}(t) - ie^{ik_o z_j}\frac{\mu(t)}{\Omega_m(t)}\hat{F}_{1m;(2)}^{j}(t)\}\delta(z-z_j), \quad (13c)$$

$$\hat{F}_{1n}^{(3)}(t,z) = N^{-1/2}\sum_{j}\{\hat{F}_{12,(\Sigma 2)}^{j}(t) - ie^{-ik_o z_j}\frac{\mu(t)}{\Omega_n(t)}\hat{F}_{1n;(2)}^{j}(t)\}\delta(z-z_j), \quad (13d)$$

here $\xi_{m,n} = Ng_{m,n}^2/(c\tilde{\gamma}_{m,n})$ and $\xi_{m,n}(\Delta_{m,n}=0) = \xi_{m,n}^o = Ng_{m,n}^2/(c\gamma_{m,n})$ are the absorption coefficients in the center of atomic transitions $|1\rangle \rightarrow |m\rangle$ and $|1\rangle \rightarrow |n\rangle$. We note that the influence of the fluctuation forces is given in Eqs. (13c) and (13d) without any approximations.

We begin our analysis of system equations Eqs. (13a) and (13b) from the time $t < t_o$ when the probe pulse field $A_l(t,z)$ has entered into the medium and $\mu(t<t_o) \cong \tilde{\gamma}_{1l}^{-1}\Omega_l^2$ (see Eq. (11b) for $\Omega_{m \neq l,n}(t<t_o) = 0$). Ignoring the relaxation for sufficiently weak atomic decoherence $\gamma_{12} << \tilde{\gamma}_{1l}^{-1}\Omega_l^2$ we find that Eq. (12a) leads to the well-known equation $(\frac{\partial}{\partial z} - ik_o)\hat{\Psi}_l = -\frac{\partial}{v_l(t)\partial t}\hat{\Psi}_l$ of the single color polariton field with the state in Eq. (4) where all the quantum correlations of the probe light pulse light are preserved during evolution [15, 16, 29, 31-33]. We note that the field propagation is described by the solution $\hat{\Psi}_l(t,z) = e^{ik_o z}\hat{\Psi}_{l;o}(z - \int^t v_l(t')dt')$ with a simple linear dispersion relation $\omega = v_l(t)k$. Therefore in the Schrödinger pictures, the initial state $|\varphi_l\rangle|1\rangle_{atoms}$ is transformed unitary to the pure dark-state of the polariton field $\hat{\Psi}_l(t,z)$. In particular for initial coherent state of the probe pulse, we have

$$|\varphi_l\rangle = |\alpha_l\rangle = \exp\{-(1/2)|\alpha_l|^2 + \alpha_l \hat{b}_l^+\}|0\rangle_{ph}, \quad (14)$$

where $\hat{b}_l^+ = \int_{-\infty}^{\infty}dk\phi_{l;coh}(k)\hat{a}_k^+$, $\int_{-\infty}^{\infty}dk|\phi_{l;coh}(k)|^2 = 1$, (function $\phi_{l,coh}(k)$ characterizes the spectral properties of the pulse). Utilizing Eq. (14) we obtain the following coherent state of the polariton field $\hat{\Psi}_l(t<t_o,z)$

$$|\Phi_l(t_o)\rangle_\alpha = \exp\{-(1/2)|\alpha_l|^2 + \alpha_l \hat{\psi}_l^+(t_o)\}|0\rangle, \quad (15a)$$

where $|0\rangle = |0\rangle_{ph}|1\rangle_{atoms}$ is a vacuum state of the polariton field, $\hat{\psi}_l^+(t_o) = \int_{-\infty}^{\infty}dk\phi_{l;coh}(k)\hat{\psi}_{l;k}^+(t_o)$, $\hat{\psi}_{l;k}(t<t_o) = (2\pi)^{-1/2}\int_{-\infty}^{\infty}dze^{-ikz}\hat{\Psi}_{f+a}(t<t_o,z)$ (see Eq. (4)), and $|\alpha_l|^2$ being the average number of polaritons.

We can write the initial dark polariton state for $t < t_o$ for arbitrary state of the probe field only by a simple substitution of the field operators $\hat{a}_k^+$ to the operator $\hat{\psi}_{l;k}^+(t_o)$ taking into account Eq. (4). In particular we will also analyze the quantum correlations in MC-light dynamics for the well-known EPR spectrally correlated copropagating two-photon state [28, 35] where the relevant dark two-polariton state takes the form

$$|\Phi_{l,AA}(t_o)\rangle = \tfrac{1}{\sqrt{2}} \int_{-\infty}^{\infty} dk_2 \int_{-\infty}^{\infty} dk_1 \phi_{l,cor}(k_2,k_1) \hat{\psi}^+_{l;k_2}(t_o) \hat{\psi}^+_{l;k_1}(t_o) |0\rangle \quad (15b)$$

here $\langle \Phi_{l,AA}(t_o) \| \Phi_{l,AA}(t_o)\rangle = \int_{-\infty}^{\infty} dk_2 \int_{-\infty}^{\infty} dk_1 |\phi_{l,cor}(k_2,k_1)|^2 = 1$, the specific function $\phi_{l,cor}(k_2,k_1)$ is analyzed in part 4.3.

Taking into account the initial quantum states of the system equations (15a) and (15b), we will solve the system of equations (13a) and (13b) which have a symmetrical form for the quantum fields $\hat{\Psi}_m(t,z)$ and $\hat{\Psi}_n(t,z)$. In the adiabatic limit, all the equations in (13) include the same term $(1-\tfrac{\partial}{\mu \partial t})[\tfrac{1}{\mu}G(\hat{\Psi})]$ with temporal derivation of the fields which leads to the integrability of Eqs. (13a) and (13b) independent of the total number of equations $M_+ + M_-$. Using the spatial Fourier transformations in Eq. (13) $\hat{\psi}_{m(n);k}(t) = (2\pi)^{-1} \int dz e^{-ikz} \hat{\Psi}_{m(n)}(t,z)$ and $\hat{f}^{(3)}_{1,m(n)}(t,k) = (2\pi)^{-1/2} \int dz e^{-ikz} \hat{F}^{(3)}_{1,m(n)}(t,z)$ we get the simple relations between the field operators $\hat{\psi}_{m(n);k}(t)$:

$$\hat{\psi}_{m;k}(t) = \{1 + i\eta_{m;l}(k,k_o)\} \hat{\psi}_{l;k}(t) + \frac{\sqrt{n_o S}}{\mu(t)\{1+i\xi_m^{-1}(k-k_o)\}} \{\hat{f}^{(3)}_{1,l}(t,k) - \hat{f}^{(3)}_{1,m}(t,k)\}, \quad (16a)$$

$$\hat{\psi}_{n;k}(t) = \{1 + i\eta_{n;l}(k,k_o)\} \hat{\psi}_{l;k}(t) + \frac{\sqrt{n_o S}}{\mu(t)\{1-i\xi_n^{-1}(k+k_o)\}} \{\hat{f}^{(3)}_{1,l}(t,k) - \hat{f}^{(3)}_{1,n}(t,k)\}, \quad (16b)$$

where $\eta_{m;l}(k,k_o) = (\xi_l^{-1} - \xi_m^{-1})(k-k_o)/\{1+i\xi_m^{-1}(k-k_o)\}$, $\eta_{n;l}(k,k_o) = \{\xi_l^{-1}(k-k_o) + \xi_n^{-1}(k+k_o)\}/\{1-i\xi_n^{-1}(k+k_o)\}$.

Relations of Eq. (16 a) and (16 b) determines all the operators $\hat{\psi}_{m(n);k}(t)$ through the probe field operators $\hat{\psi}_{l;k}(t)$ and fluctuation forces. The relations of Eq. (16) means a reduction of the degrees of freedom of the quantum fields $\hat{\Psi}_{m(n)}(t,z)$ that reflects an important consequence of the near adiabatic multi-wave interaction, similar to the coupling of the fields $\hat{\Psi}^+_{m(n)}(t,z)$ in the dark MC states of Eq. (2). General inspection of system equations (7) and (9) shows that the new independent modes of fields $\hat{A}^+_{m(n)}(t,z)$ can be excited only at the nonadiabatic interactions, where the field-atoms dynamics becomes much more complicated (this situation in some particular cases will be studied elsewhere). We note that the temporal behavior of the field operators $\hat{\psi}_{m(n);k}(t)$ and $\hat{\psi}_{l;k}(t)$ in Eqs. (16a) and (16b) is independent of the Rabi frequencies $\Omega_{m,n}(t)$ and decay constant $\gamma_2$. This very universal coupling between the quantum fields is destroyed only by the difference in the fluctuation forces $\sim \hat{f}^{(3)}_{m(n)-l}(t,k)$ on different optical transitions. Based on Eqs. (16a) and (16b), we find the following relations between the field operators in space-time domain:

$$\hat{\Psi}_{m,(n)}(t,z) = \int_0^L dz' f_{m(n);l}(z-z')\hat{\Psi}_l(t,z') + \delta\hat{\Psi}_{st,m(n);l}(t,z), \tag{17a}$$

$$\delta\hat{\Psi}_{st;m;l}(t,z) = \sqrt{N}\xi_m\mu^{-1}(t)\int_0^z dz'\exp\{(ik_o - \xi_m)(z-z')\}\{\hat{F}^{(3)}_{1,l}(t,z') - \hat{F}^{(3)}_{1,m}(t,z')\}, \tag{17b}$$

$$\delta\hat{\Psi}_{st;n;l}(t,z) = \sqrt{N}\xi_n\mu^{-1}(t)\int_z^L dz'\exp\{(ik_o + \xi_n)(z-z')\}\{\hat{F}^{(3)}_{1,l}(t,z') - \hat{F}^{(3)}_{1,n}(t,z')\}. \tag{17c}$$

The functions $f_{m(n),l}(z-z')$ are analyzed in Appendix 1. Eq. (17a) means that all the light fields $\hat{A}^+_{m(n)}(t,z)$ propagate together in the medium with united group velocity v at the same time each field $\hat{\Psi}_{m(n)}(t,z)$ gets its own quantum noise if $\delta\hat{\Psi}_{sl;m,l}(t,z) \neq 0$ and $\delta\hat{\Psi}_{st;n,l}(t,z) \neq 0$. Thus all the quantum fields $\hat{\Psi}_{m(n)}(t,z)$ are coupled to each other in the medium.

Applying Eqs. (16a) and (16b) to Eqs. (13a) and (13b) using algebraic calculations we find an ordinary differential equation for $\hat{\psi}_{l;k}(t)$ and obtain its analytical solution:

$$\hat{\psi}_l(t,k) = \hat{\psi}_{det;l}(t,k) + \delta\hat{\psi}_{st;l}(t,k), \tag{18a}$$

where $\hat{\psi}_{det;l}(t,k)$ describes deterministic evolution of the initial quantum field $\hat{\psi}_l(t,k)$ due its coupling with new quantum fields $\hat{\psi}_m(t,k)$ and $\hat{\psi}_n(t,k)$:

$$\hat{\psi}_{det;l}(t,k) = T(t-t_o,k,k_o)\hat{\psi}_l(t_o,k). \tag{18b}$$

The function $\delta\hat{\psi}_{st;l}(t,k)$ resulted from the quantum noise influence (see Appendix 2) is

$$\delta\hat{\psi}_{st;l}(t,k) = \sqrt{n_o S}\int_{t_o}^t dt'\mu(t')\frac{T(t-t',k,k_o)}{I(t',k,k_o)}\hat{f}^{(4)}_{1,l}(t',k), \tag{18c}$$

here we have also introduced the following functions

$$T(t-t_o,k,k_o) = \frac{I(t_o,k,k_o)}{I(t,k,k_o)}\exp\{-i\int_{t_o}^t dt'\tilde{\omega}(t',k,k_o)\}, \tag{19a}$$

$$\tilde{\omega}(t,k,k_o) = \{\frac{-i\gamma_{12} + \mu(t)\xi_l^{-1}(k-k_o) - \beta(t,k,k_o)}{I(t,k,k_o)}\}, \tag{19b}$$

$$I(t,k,k_o) = 1 - (\gamma_{12}/\mu(t)) + i\beta(t,k,k_o)/\mu(t), \tag{19c}$$

$$\beta(t;k,k_o) = \sum_m^{M_+}\Gamma_m(t)\eta_{m;l}(k,k_o) + \sum_n^{M_-}\Gamma_n(t)\eta_{n;l}(k,k_o), \tag{19d}$$

where $\beta(t \leq t_o;k,k_o) = 0$ before switching on the new control laser fields. The final solution takes the following form:

$$\hat{\Psi}_{m,(n)}(t,z) = \hat{\Psi}_{det;m(n)}(t,z) + \delta\hat{\Psi}_{st;m(n)}(t,z), \tag{20a}$$

$$\hat{\Psi}_{det;l}(t,z) = (2\pi)^{-1/2} \int_{-\infty}^{\infty} dk e^{ikz} \hat{\psi}_{det;l}(t,k), \tag{20b}$$

$$\delta\hat{\Psi}_{st;l}(t,z) = (2\pi)^{-1/2} \int_{-\infty}^{\infty} dk e^{ikz} \delta\hat{\psi}_{st;l}(t,k), \tag{20c}$$

$$\hat{\Psi}_{det;m,(n)}(t,z) = \int_{0}^{L} dz' f_{m,(n);l}(z-z') \hat{\Psi}_{det;l}(t,z'), \tag{20d}$$

$$\delta\hat{\Psi}_{st;m,(n)}(t,z) = \int_{0}^{L} dz' f_{m,(n);l}(z-z') \delta\hat{\Psi}_{st;l}(t,z') + \delta\hat{\Psi}_{st,m(n);l}(t,z), \tag{20e}$$

where as seen from Eqs. (17b) and (17c) $\delta\hat{\Psi}_{st,l;l}(t,z) = 0$.

Thus we note a remarkable property in the system equations (13a)-(13b) which leads to the unique solution in Eqs. (17a) and (20) independent of the total number $M_+ + M_-$ of the coupled quantum light fields. Here we should note that the adiabatic conditions dramatically change the evolution of each optical field $\hat{A}^+_{m(n)}(t,z)$ which leads to one united field characterized by the one dispersion relation in Eq. (19b) (see below). Such multi-waves quantum evolution opens a convenient method for perfect control of the MC-light properties which is studied in detail in the following sections.

The most interesting deterministic term of the adiabatic evolution $\hat{\psi}_{det;l}(t,k)$ is characterized by the two main factors in Eq. (19a). Temporal dependence of the first pre-exponential factor gives the purely adiabatic reversible evolution of the fields (see also Appendix 2) whereas the exponential factor determines the wave dynamics of coupled fields which also has irreversible dissipation terms. In the case of a negligible small splitting between the two lowest levels: $k_o \xi_l^{-1} = \omega_{21}/(c\xi_l) \ll 1$, $k_o \xi_m^{-1} \ll 1$ we obtain from Eq.(17a) by using Eq. (A6) and (A7) (see Appendix 1):

$$\hat{\Psi}_{det;m}(t,z)\Big|_{\xi_l l_o \gg 1, k_o \xi_l^{-1} \ll 1} \cong \xi_m \int_{0}^{z} dz' \exp\{-\xi_m(z-z')\} \hat{\Psi}_{det;l}(t,z'+\tfrac{1}{\xi_l}), \tag{21a}$$

$$\hat{\Psi}_{det;n}(t,z)\Big|_{\xi_l l_o \gg 1, k_o \xi_l^{-1} \ll 1} \cong \xi_n \int_{z}^{L} dz' \exp\{\xi_n(z-z')\} \hat{\Psi}_{det;l}(t,z'+\tfrac{1}{\xi_l}). \tag{21b}$$

Thus the m-th field operator $\hat{\Psi}_{det;m}(t,z)$ is determined by the operator of probe field $\hat{\Psi}_{det;l}(t,z')$ through the nonlocal asymmetric spatial coupling of the fields with the coordinate relation $z' < z + 1/\xi_l$. Whereas the field $\hat{\Psi}_{det;n}(t,z)$ is coupled with field $\hat{\Psi}_{det;l}(t,z')$ through nonlocal spatial coupling with the opposite coordinate relation $z' > z + 1/\xi_l$, which is a result of the interaction between the field $E_l$ with the fields $E_m$ and $E_n$ propagating in the opposite direction. The similar effect takes place for fluctuation forces $\delta\Psi_{st;m(n),l}(t,z)$ (see (17b) and (17c)). Taking into account the initial quantum state of the probe pulse we see from Eqs. (21a) and (21b) that in the optically dense media $l_o \xi_{m,n} \gg 1$ ($l_o$ is a initial spatial size of the probe light pulse $E_l$) the new fields $\hat{\Psi}_m(t,z)$ and $\hat{\Psi}_n(t,z)$ will be quantum copies of the original probe field $\hat{\Psi}_l(t,z)$ if the spatial correlated length $l_{cor}$ of the initial field is large enough $l_{cor}\xi_{m,n} \gg 1$. In accordance with Eqs. (21a) and (21b) and Eq. (20) a quality of the quantum copying processes and transference of all the quantum correlations to the new fields will be determined by the parameters of the deterministic and chaotic parts of the field $\hat{\Psi}_l(t,z)$ during the interaction time with the new fields $\hat{\Psi}_{m(n)}(t,z)$. In the next section we analyze the details of quantum dynamics and control of the new fields generation in the most important temporal and spatial aspects.

### III. Control of the MC field Dynamics

Below we analyze the deterministic part of the quantum evolution of the MC-light taking into account the physical conditions of the typical experiments with a slow light. Usually we have a weak enough relaxation between the two long-lived levels (in our case – levels 1 and 2 ) so assuming $\gamma_{12} \ll \gamma_{1m}, \gamma_{1n}, \mu(t)$, we obtain the following expression from Eqs. (19a) and (19b):

$$\hat{\psi}_{det'm(n),k}(t) = T_{m(n);l}(t-t_o;k,k_o)\hat{\psi}_{l;k}(t_o) \cong J_{m(n);l}(t;k,k_o)\exp\{-\gamma_2(t-t_o) - i\int_{t_o}^{t} dt'\omega(t';k,k_o)\}\hat{\psi}_{l;k}(t_o), \qquad (22a)$$

where a dispersion relation $\omega(t;k,k_o)$ of the coupled light is determined by the relation:

$$\omega(t;k,k_o) = \frac{(k-k_o)\mu(t)/\xi_l - \beta(t;k,k_o)}{\{1+i\beta(t;k,k_o)/\mu(t)\}}, \qquad (22b)$$

and

$$J_{m(n);l}(t;k,k_o) = \frac{\{1+i\eta_{m(n);l}(k,k_o)\}}{\{1+i\beta(t;k,k_o)/\mu(t)\}}. \qquad (22c)$$

where $J_{m(n);l}(t>t_1;k,k_o) = 1$ if only the one control field $\Omega_{m'(or\,n')}(t) \neq 0$ is switched on.
In the second step we again take into account the small energy splitting between the two lowest atomic levels so that $k_o/\xi_{m,n} \ll 1$ and $|k_o| \ll |k|$ therefore we get

$$\eta_{m;l}(k,k_o)\big|_{k_o=0} = \eta_{m;l}(k) = k(\xi_l^{-1} - \xi_m^{-1})(1+i\xi_m^{-1}k)^{-1}, \qquad (23a)$$

$$\eta_{n;l}(k,k_o)\big|_{k_o=0} = \eta_{n;l}(k) = k(\xi_l^{-1} + \xi_n^{-1})(1-i\xi_n^{-1}k)^{-1}. \qquad (23b)$$

The relation $k_o/\xi_{m,n} \ll 1$ facilitates the phase matching fulfillment for the interaction between the light fields, which is more easily realized in the optical dense medium where $l_o\xi_l \gg 1$. These conditions lead to the inequalities $\eta_{m(n);l}(k) \ll 1$ and similarly to the relation $|\beta(t;k,0)| \ll |\mu(t)|$ for all the spectral components of the light fields. Using Eqs. (20a) - (20e) and the inequalities in Eq. (15b) and Eq. (19a) we get $J_{m(n);l}(t;k,k_o = 0) \approx \exp\{i\theta_{m(n);l}(t,t_o;k)\}$ (where $\theta_{m(n)}(t,t_o;k)$ see Appendix 2) and the following decomposition for the dispersion relation of Eq. (22b):

$$\omega(t;k,k_o = 0) = v(t)k - i\delta\omega_{kk}^{"}(t)k^2/2 + ...,  \quad (24a)$$

$$v(t) = \omega_k^{'} = \sum_m^{M_+}\Gamma_m\xi_m^{-1} - \sum_n^{M_-}\Gamma_n\xi_n^{-1} = c\{\sum_m^{M_+}\Omega_{m;o}^2/(Ng_m^2) - \sum_n^{M_-}\Omega_{n;o}^2/(Ng_n^2)\} = \sum_m^{M_+}v_m - \sum_n^{M_-}v_n, \quad (24b)$$

$$\delta\omega_{kk}^{"} = i\omega_{kk}^{"} = 2\{\sum_{m,m'}^{M_+}\Gamma_m\Gamma_{m'}(\xi_m^{-1} - \xi_{m'}^{-1})^2 + \sum_{n,n'}^{M_-}\Gamma_n\Gamma_{n'}(\xi_n^{-1} - \xi_{n'}^{-1})^2 + \sum_{m,n}^{M_+,M_-}\Gamma_m\Gamma_n(\xi_m^{-1} + \xi_n^{-1})^2\}/\mu, \quad (24c)$$

where $v(t)$ is a united group velocity of all the interacting fields, the term $-i\delta\omega_{kk}^{"}$ is a second order dispersion determining a spatial broadening of the MC-light pulse.

As seen from Eq. (24b) the group velocity $v(t)$ does not depend on the spectral detunings $\Delta_m$ and $\Delta_n$ and the velocity $v(t)$ can be controlled only by the manipulation of the laser field amplitudes $\Omega_{m,n}(t)$. Let us consider the possible regimes of the MC-light generation.

### III.a. Traveling MC-field

If $M_- = 0$ and $M_+ > 1$ we have the traveling MC-light field with the following group velocity

$$v(t) = v_{travel}(t) = \sum_m^{M_+} v_m(t). \quad (25)$$

where the spatial dispersion will be

$$\delta\omega_{kk;travel}^{"} = 2\{\sum_{m,m'}^{M_+}\Gamma_m\Gamma_{m'}(\xi_m^{-1} - \xi_{m'}^{-1})^2\}/\mu. \quad (26a)$$

The dispersion of traveling MC-light field becomes minimum for spectral detunings

$$\delta\omega_{kk;travel}^{"} = 2\{\sum_{m,m'}^{M_+}\Gamma_m\Gamma_{m'}(1/\xi_m^o - 1/\xi_{m'}^o)^2\}/\mu, \quad (26b).$$

In this case the dispersion can be even equals to zero if $\gamma_m/g_m^2 = Const$.

### III.b. Stationary MC-field

Most interesting case of quantum manipulations of MC-field evolution takes place at the complete stopping point $v(t) = 0$ of the light fields envelopes:

$$\sum_m^{M_+}\Omega_{m;o}^2(t)/g_m^2 - \sum_n^{M_-}\Omega_{n;o}^2(t)/g_n^2 = 0. \quad (27)$$

The relation generalizes the recent results obtained for the two- and three-color lights in [27]. The stopping is a result of the strong interaction between all the coupled fields propagating in opposite directions. While keeping the condition (27) it is possible to vary the amplitudes of different control laser fields thereby changing the electromagnetic field amplitudes of the different components in accordance with relations $\hat{A}_m(t,z)/\hat{A}_n(t,z) \cong \Omega_m(t)g_n/(\Omega_n(t)g_m)$. Therefore a number of control regimes the value and direction of the MC-light velocity can be realized by the manipulation of laser field's amplitudes. Such manipulations of group velocity and electric field amplitudes of the MC-light fields look promising for controlling the spectrally selective weak interactions of the chosen frequency components of light fields with different resonant atomic transitions of media or even with selected single atoms. This issue will be studied elsewhere.

Maximum time of the MC stationary light control is limited by both the relaxation constant $\gamma_{12}$ and the irreversible spatial spreading of the MC-light envelope which are determined by the dispersion term $\delta\omega_{kk}^{''}$ of Eq. (24c). Putting Eq. (27) in Eq. (24c) we get to the following formula for the second order dispersion:

$$\delta\omega_{kk}^{''}\big|_{v=0} = \delta\omega_{kk;l}^{''}(t) = 2\{\sum_{m}^{M_+} v_m(t)\xi_m^{-1} + \sum_{n}^{M_-} v_n(t)\xi_n^{-1}\}. \tag{28a}$$

Taking into account Eq. (27) in Eq. (28) we can noticeably minimize the dispersion effect satisfying the conditions: $\text{Im}(\xi_m^{-1} - \xi_l^{-1}) = 0$ and $\text{Im}(\xi_n^{-1} + \xi_l^{-1}) = 0$ at the following spectral condition

$$\sum_{m}^{M_+} v_m(t)(\Delta_m/g_m^2) + \sum_{n}^{M_-} v_n(t)(\Delta_n/g_n^2) = 0 \tag{29a}$$

with the following particular solution related to the initial probe field $E_l$:

$$\Delta_m/g_m^2 - \Delta_{l'}/g_l^2 = 0, \quad \Delta_n/g_n^2 + \Delta_{l'}/g_{l'}^2 = 0, \tag{29b}$$

Thus in accordance with Eq.(29b) frequency signs are the same for the detunings $\Delta_m$ and $\Delta_{l'}$ whereas the signs are the opposite for $\Delta_n$ and $\Delta_{l'}$ (for the counterpropagating fields). Eq. (29) gives a universal condition of the minimal spreading which is independent of the total number of the fields $M_+$ and $M_-$. The condition generalizes the results for stationary two- and three-color lights obtained in the works [27]. Using Eqs. (28) or (29) in Eq. (26) we finally obtain a simple important relation for the minimum dispersion of the stationary MC-light:

$$\delta\omega_{kk}^{''}\big|_{v=0} \equiv \delta\omega_{kk;o}^{''}(t) = 2\{\sum_{m}^{M_+} v_m(t)/\xi_m^o + \sum_{n}^{M_-} v_n(t)/\xi_n^o\}, \tag{28b}$$

Here we should note that the dispersion term of (28b), after putting in (24a), gives a spectrally selective absorption of the united k-th mode of MC-field leading to the irreversible loss in quantum correlations and energy of the MC-field. The increase of the number of coupled light fields $M_+ + M_-$ enhances the spatial broadening of stationary MC-field envelopes. However using of the atomic media, with large resonant absorption coefficients $\xi_m^o$ and $\xi_n^o$, provides the possibility for increasing the number of coupled fields $M_+ + M_-$ along with enhanced stationary light amplitudes. Temporal and spatial effects of the spectrally selective absorption of the quantum MC light fields are studied in details below for particular quantum states of light, taking into account the optimal spectral conditions obtained above.

### IV. Dynamics of the Quantum MC-fields

The properties of quantum fields can be analyzed using Eqs. (18a) and (18b) while the quantum average of the electromagnetic fields are given

$$\langle \hat{A}_{m,n}(t,z) \rangle = \Omega_{m,n}(t)(Ng_{m,n}^2)^{-1/2} \langle \hat{\Psi}_{m,n}(t,z) \rangle, \tag{30a}$$

$$\langle \hat{A}_{p'}^+(t_2,z_2)\hat{A}_p(t_1,z_1) \rangle = \Omega_{p'}^*(t_2)\Omega_p(t_1)(Ng_{p'}^2 Ng_p^2)^{-1/2} \langle \hat{\Psi}_{p'}^+(t_2,z_2)\hat{\Psi}_p(t_1,z_1) \rangle, \tag{30b}$$

where:

$$\langle \hat{\Psi}_{m,n}(t,z) \rangle = G_{m,n;l}(z-z') \otimes \langle \hat{\Psi}_{det;l}(t,z') \rangle, \tag{31a}$$

$$\langle \hat{\Psi}_l(t,z') \rangle = \langle \hat{\Psi}_{det;l}(t,z') \rangle = (2\pi)^{-1/2} \int_{-\infty}^{\infty} dk e^{ikz'} U_{k,k_o}(t-t_o) \langle \hat{\psi}_{l;k}(t_o) \rangle, \tag{31b}$$

$$\langle \hat{\Psi}_{p'}^+(t_2,z_2)\hat{\Psi}_p(t_1,z_1) \rangle = G_{p';l}^+(z_2-z') \otimes G_{p;l}(z_1-z) \otimes \langle \hat{\Psi}_{det;l}^+(t_2,z')\hat{\Psi}_{det;l}(t_1,z) \rangle + \langle \delta\hat{\Psi}_{st;p'}^+(t,z_1)\delta\hat{\Psi}_{st;p}(t,z_2) \rangle, \tag{31c}$$

$$\langle \hat{\Psi}_{det;l}^+(t_2,z')\hat{\Psi}_{det;l}(t_1,z) \rangle = (2\pi)^{-1} \int\int_{-\infty}^{\infty} dk_2 dk_1 e^{ik_2 z' + ik_1 z} U_{k_2,k_o}(t_2-t_o) U_{k_1,k_o}(t_1-t_o) \langle \hat{\psi}_{l;k_2}^+(t_o)\hat{\psi}_{l;k_1}(t_o) \rangle, \tag{31d}$$

where $\langle ... \rangle$ is a quantum average over the initial quantum states (below we use the states $|\Phi_l(t_o)\rangle_\alpha$ and $|\Phi_l(t_o)\rangle_{cor}$ of Eqs. (15a) and (15б)); $\langle \delta\hat{\Psi}_{st;p'}^+(t,z_1)\delta\hat{\Psi}_{st;p}(t,z_2) \rangle$ are calculated using Eqs. (17b), (17c) and Eqs. (21c), (21e) and the quantum correlators of fluctuation forces in Eqs (8b)-(8d);

For simplicity, we impose the following relations on the fluctuation parts of the field operators $\langle \delta\hat{\Psi}_{st;p'}^+(t,z_1)\delta\hat{\Psi}_{st;p}(t,z_2) \rangle = \langle \delta\hat{\Psi}_{st;p'}(t,z_1)\delta\hat{\Psi}_{st;p}(t,z_2) \rangle = 0$ which corresponds to the vacuum state of quantum noise bath. The formulas for higher quantum correlators can be written similar to Eqs. (30) - (31) through the operators of initial probe fields $\hat{\Psi}_l(t,z)$ and $\hat{\Psi}_l^+(t,z')$. The basic properties of spatial and temporal control of MC-fields are more easily analyzed using the coherent state of probe light (15a).

### IV.a. Coherent State of MC-fields

The dynamics of the coherent state often demonstrates the classical properties of light and it takes place here for $\langle \hat{\Psi}_l(t_2,z_2) \hat{\Psi}_l(t_1,z_1) \rangle_\alpha = \langle \hat{\Psi}_l(t_2,z_2) \rangle_\alpha \langle \hat{\Psi}_l(t_1,z_1) \rangle_\alpha$, so we can analyze only the first order correlator using Eqs. (31a) and (31b). For convenience we assume a Gaussian spectral shape of the probe pulse field:

$$\phi_{l,coh}(k) = (l_o/\sqrt{\pi})^{1/2} \exp\{-\tfrac{1}{2}(kl_o)^2 + ik(z_o - v_l t_o) + i(\vartheta_l - \phi_l)\},\tag{32}$$

where $z_o$ is a coordinate of the probe field $A_l$ at $t=t_o$, $\vartheta_l$ is a constant phase shift of the field, $l_o = v_l \delta t_o$ is the initial spatial size of the probe pulse in the medium, which is determined by the linear dispersion relation $\omega = v_l k$ of the probe light field for $t<t_o$ (see Eqs. (24)-(26)). Applying Eq. (32) in Eqs. (30 a) and (31 b) and putting $U_{k,k_o}(0)=1$ we get the electric field amplitude of the probe field before the new control fields $\Omega_{m,n}(t)$ is switched on:

$$\langle \hat{A}_l(t_o,z) \rangle_\alpha = A_{l;o} \exp\{i\vartheta_l - \tfrac{1}{2}(z+z_o - v_l t_o)^2 / l_o^2\}.\tag{33}$$

Solution (33) describes a light pulse propagating with the constant slow group velocity $v_l$ and with the envelope amplitude $A_{l,o} = (\sqrt{\pi}c\delta t_o)^{-1/2} a_l$ which is independent of the constant group velocity ($|a_l|^2$ is the average number of polaritons). Using the solution in Eqs. (19b) and (21b) for switching on the new control laser fields ($\Omega_{m,n}(t>t_o) \neq 0$) and then performing the calculations similar to Eq. (33) and using Eqs. (3.1) and (3.2) of Appendix 3 we obtain the following quantum average for the field $\langle \hat{\Psi}_l(t,z) \rangle_\alpha$:

$$\langle \hat{\Psi}_l(t,z) \rangle_\alpha = \sqrt{c/v_l(t_o)} \frac{l_o}{l_l(t)} A_{l;o} \exp\{-\gamma_2(t-t_o) + i(\vartheta_l - \phi_l)\} \exp\{-\tfrac{1}{2}[z+z_o - v_l t_o - \int_{t_o}^t v(t')dt' + \delta z_l(t)]^2 / l_l^2(t)\},\tag{34}$$

where the group velocity $v(t)$ is given in Eq.(24b) and,

$$l_{m(n)}(t) = \sqrt{l_o^2 - \delta l_{m(n)}^2(t) + \int_{t_o}^t \delta\omega_{kk}''(t')dt'}\tag{35}$$

is a new spatial size of the pulse envelope; $\delta z_l(t)$ and $\delta l_{m(n)}(t)$ are small spatial shifts (see Appendix 3). After transference to a single color polariton field we find in Eq. (35) $l_{m(n)}(t) \cong l(t) = \sqrt{l_o^2 + \int_{t_o}^{t_o+\tau} \delta\omega_{kk}''(t')dt'}$ (where $\tau$ is the manipulation time of the MC-field). In the optically dense media we have from Eq. (34), the l-th and m-th field envelopes are shifted to each other at the distance $\delta z_{ml} = |\delta z_m - \delta z_l| \approx |\xi_m^{-1} - \xi_l^{-1}|$ and at the distance $\delta z_{nl} \approx |\xi_n^{-1} + \xi_l^{-1}|$ between the two l-th and n-th fields. We note the correlation between spatial shifts of field envelopes and second order dispersion $\delta\omega_{kk}''(t')$ for the simplest case of the two-color stationary field (where $\delta\omega_{kk}''(t')$ is determined by Eq. (26b) for the traveling MC-field and Eq. (28b) - for the stationary light $v(t) = 0$). We note that the spatial spreading decreases at the optimal spectral conditions in Eqs. (26b) and (28b). In accordance with Eqs. (18a) and (18b) the quantum averages of the field operators satisfy $\langle \hat{\Psi}_m(t_2,z_2) \rangle_\alpha \approx \langle \hat{\Psi}_l(t_1,z_1) \rangle_\alpha$ in the optically dense medium. Using Eqs. (31c)-(31d) we find the two particle correlator for the coherent polariton state (15a) $\langle \hat{\Psi}_m(t_2,z_2)\hat{\Psi}_n(t_1,z_1) \rangle_\alpha \approx \langle \hat{\Psi}_l(t_2,z_2) \rangle_\alpha \langle \hat{\Psi}_l(t_1,z_1) \rangle_\alpha$ and for higher order quantum correlators. Assuming that the MC-field is localized completely in the medium we obtain for the total number of polaritons $N(t)$ of the field:

$$N(t) = \int dk \langle \hat{\psi}^+_{l,k}(t)\hat{\psi}_{l,k}(t) \rangle_\alpha = \int_0^L dz \langle \hat{\Psi}^+_l(t,z)\hat{\Psi}_l(t,z) \rangle_\alpha = \frac{l_o}{l_l(t)}|a_l|^2 \exp\{-2\gamma_2(t-t_o)\}. \quad (36)$$

The spatial spreading of the MC-light envelope is accompanied by the energy loss of the diffusion type $\sim l_o/l_l(t)$. Typical spreading time $\delta t_{spread}$ of complete loss is found from Eq. (35) and (29) being comparable and larger than:

$$\delta t_{spread} \approx l_o^2 / \delta\omega''_{kk}(v=0) = l_o^2 / \{2\sum_m^{M_+} v_m / \xi_m^o + 2\sum_n^{M_-} v_n / \xi_n^o\}. \quad (37)$$

Physical reason of the MC-polarion field absorption given in Eq. (36) and spatial broadening of the field envelope in Eq. (35) can be understood by taking into account the comparison of field envelopes in Eq. (34) and $\delta z_{m(n)l} \neq 0$ with the envelopes in the pure dark MC-polariton fields given in Eq. (6). The stationary MC-field in Eq. (34) differs from the pure dark MC state due to the spatial shifts of the field envelopes. Therefore we conclude that the spatial field shifts result in absorption and spatial spreading of the MC-field which leads to the decrease of $\delta z_{n(m)l}/l(t) \to 0$ and moving thereby nearer to the pure MC-dark state. We stress that the traveling MC-field can be originally very near to the pure dark state due to the possibility of $\delta z_{ml} \to 0$ while the dissipation processes for stationary MC-field becomes minimized in the optically dense media only if the initial spatial envelope is large enough $l_o >> \xi_l^{-1} + \xi_n^{-1}$.

Increase of the light field number $M_+ + M_-$ opens the possibility of intensive swapping of the information to the many frequency components of the MC-light whereas large fields number decreases the manipulation time due to the shorter spreading time $\delta t_{spread}$. Using Eq.(30 a) and (34) we find the electric field amplitudes $A_{m,n}$ of the stationary MC-light ($v(t) = 0$):

$$\left\langle \hat{A}_{m(n)}(t,z) \right\rangle_\alpha = \frac{\Omega_{m(n)}(t) g_l}{\Omega_l(t_o) g_{m(n)}} \frac{l_o}{l_{m(n)}(t)} A_{l;o} \exp\left\{-\gamma_2(t-t_o) - \frac{[z+z_o - v_l t_o + \delta z_{m(n)}]^2}{2 l_{m(n)}^2(t)}\right\} e^{i(\vartheta_l + \phi_{m(n)} - \phi_l)}. \tag{38}$$

As seen from Eq. (38) the amplitude $A_{m,n}$ can be considerably enhanced even in comparison with the initial probe pulse amplitude $A_{l,o}$ if $\Omega_{m,n}(t)/\Omega_l(t_o) \gg 1$. Such enhanced field can stay in the medium until the $\gamma_{12}$ relaxation and large spatial broadening ($l_o/l_{m(n)}(t) \to 0$) destroy the coherent part of the MC-light field. The manipulation of the stationary light amplitudes looks promising for the enhancement of the weak photon-photon interactions and the lengthening of the interaction time however the spatial broadening problem becomes especially important for large field amplitudes $A_{m,n}$ of the stationary field. This issue will be studied in detail elsewhere taking into account possible schemes of the quantum nondemolition control based on the EIT [25].

### IV.b. Coherent State. MC-wavelength Conversion

It is possible to switch the stationary MC-light into any single traveling light field $A_m$ with $m \in [1,...,M_+]$ by manipulating the amplitudes of the control laser fields $\Omega_m(t>t_1) \neq 0$ $\Omega_{l \neq m}(t>t_1) = 0$ or to the light field $A_n$ $n \in [1,...,M_-]$ if $\Omega_n(t>t_1) \neq 0$ $\Omega_{l \neq n}(t>t_1) = 0$). It is also possible to generate some specific superposition of the number copropagating quantum fields $A_m$ or $A_n$ (see part 3.1). Using Eq. (25) for the traveling group velocity $v_{travel}(t)$ and Eqs. (30 a) and (34) we find the amplitude of the m-th component in the MC-light at time $t_{out} > t_1$ on the medium output (z=L)

$$\left\langle A_{m;o}(t) \right\rangle_\alpha \cong \sqrt{\frac{v_m(t_{out})}{v_l(t_o)}} \frac{l_o}{l_m(t_{out})} A_{l;o} \exp\{-\gamma_2(t-t_o) + i(\vartheta_l + \phi_m - \phi_l)\} \exp\{-\tfrac{1}{2}(t-t_{out})^2/(\delta t_{m,out})\}, \tag{39a}$$

with the temporal duration $\delta t_{m,travel} = l_m(t_{out})/v_{travel}(t_{out})$.

Since $v_{travel}(t)$ increases with the number of the coupled fields, the MC-light field will have a shorter temporal duration $\delta t_{m,travel}$ (at the same Rabi frequencies of the control fields) therefore the MC-light will be irradiated to the free space from the medium with the new spatial size $l_{MC;out} = cl(t_{out})/v_{travel}(t_{out})$. Using Eq. (34) we find an energy $W_m$ of the irradiated in the m-th field component:

$$W_m(t>t_{out}) = W_{l,o} \frac{\omega_m}{\omega_l} \frac{v_m(t_{out})}{v_{travel}(t_{out})} \frac{l_o}{l(t_{out})} \exp\{-2\gamma_2(t_{out}-t_o)\}, \tag{39b}$$

and the total energy $W_{out}$ of the MC-light will be

$$W_{out} = \sum_m W_m = W_{l,o} \frac{\overline{\omega}}{\omega_l} \frac{l_o}{l(t_{out})} \exp\{-2\gamma_2(t_{out}-t_o)\}, \tag{40}$$

where $W_{l,o}$ is an energy of the initial probe pulse, $\bar{\omega} = \sum_m \omega_m v_m(t_{out})/v_{travel}(t_{out})$ is an average frequency of the MC-light fields. We note that the light fields $A_n(t,z=L)=0$ in accordance with Eq. (18b) and the situation similar to Eq. (39) and (40) takes place for the irradiated fields $A_n(t,z=0)$ whereas $A_m(t,z=0)=0$ (see Eq.(18a)). We note that though the m-th light amplitude $A_m$ can be larger than the initial probe pulse amplitude $A_{l,o}$ but the time duration of the output m-th light field pulse becomes shorter in the relevant proportion so the average number of photons in the MC-light is limited by the photon number of the initial probe field. Thus it is possible to amplify the output energy of the output MC-light field by increasing the average frequency $\bar{\omega}/\omega_l >> l(t_{out})/l_o(t_o)$ and satisfying weak $\gamma_{12}$ relaxation.

We note that the spectrally selective disturbance and absorption of the slow light field, effect negatively on the quantum entanglement of slow light field [32, 33]. In the next section we demonstrate the influence of the similar disturbances in the quantum evolution of stationary MC-fields on the fragile spatial quantum correlations for the case of the two-photon spectrally correlated probe light. Such correlations are especially important for quantum communication protocols and quantum imaging processes based on the spectrally correlated fields.

**IV.c. Two-photon Entangled State: Quantum Correlations in the MC-light**

We note that a maximum possible spectral width of the slow light is determined by the spectral window transparency of EIT. Principally the spectral width can reach a GHz range for semiconductor materials [13]. Let us assume that the two photon field in the quantum EPR state is launched into the medium creating a spectrally correlated two-polariton field in Eq. (15b). Such narrow spectral correlations can be principally realized at the coherent Raman scheme of two-photon field generation using the protocol of Duan, Lukin, Cirac, and Zoller [34]. Without loss of essential physics of the quantum correlation we use the following form of the initial wave function [28]:

$$\phi_{l,cor}(k_2, k_1) = \Gamma_1(k_2 + k_1)\Gamma_2(k_1)\Gamma_2(k_2). \tag{41}$$

with the Gaussian functions for $\Gamma_1(k_1 + k_2)$ and $\Gamma_2(k_1)$, where $\Gamma_2(k) = (b/\sqrt{\pi})^{1/4} \exp\{-\frac{1}{2}b^2 k^2\}$ describes the spectral distribution of the $k_1$ and $k_2$ polaritons (where $b^{-1} = \gamma_{ph}/v_l$ is the k-spectrum width, $\gamma_{ph}$ is the larger parameter, which determines the spectrum width of the probe light field; $\Gamma_1(k_2 + k_1) = [(2a^2 + b^2)/\pi]^{1/4} \exp\{-\frac{1}{2}a^2(k_2 + k_1)^2\}$ describes spectral correlations between the two polaritons with correlation length $a^{-1} = \gamma_{cor}/v_l$ in a k-spectrum; Maximum of $\Gamma_1(k_2 + k_1)$ at

$k_2 + k_1 = 0$ corresponds to the energy conservation $\omega_1 + \omega_2 = \omega_p$ at the two photon down conversion with spectral width $\gamma_{cor}$ of the photons correlation. Spectral width $\gamma_{cor}$ is assumed to be narrow compare to the spectral width of a each single photon field $\gamma_{ph}$: $\gamma_{cor} \ll \gamma_{ph}$ ($b \ll a$). Similar to parts 4.1-4.2 we taking into account the solution in Eqs. (18)-(20) at the optimal dispersion relation (21a) and following to the analysis of the two-photon wave function given in [35] we introduce a *two-polariton wave function* $\Psi_{ll}(t',z';t,z;\tau)$

$$\Psi_{ll}(t',z';t,z;\tau) = \frac{1}{\sqrt{2}}\langle 0|\hat{\Psi}_l(t',z';\tau)\hat{\Psi}_l(t,z;\tau)|\Phi_{l,cor}(t_o;2)\rangle = N_2^{1/2}(\tau)Y_{norm}(Z_1,Z_1';a,b_l(\tau)), \tag{42}$$

where taking into account $\langle 0|\hat{\Psi}_{det;m'(n')}(t,z)\delta\hat{\Psi}_{st;m(n)}(t,z)|\Phi_{l,cor}(t_o;2)\rangle = \langle 0|\delta\hat{\Psi}_{st;m'(n')}(t,z)\delta\hat{\Psi}_{st;m(n)}(t,z)|\Phi_{l,cor}(t_o;2)\rangle = 0$ we find

$$Y_{norm}(Z_1 Z_2;a,b_l(\tau)) = \frac{1}{\sqrt{\pi a l_{coh,(2)}(\tau)}} \exp\{-\frac{Z_1^2 + Z_2^2}{2l_{l(2)}^2(\tau)}\} \exp\{-\frac{[Z_2 - Z_1]^2}{2l_{coh,(2)}^2(\tau)}\}, \tag{43a}$$

$$N_2(\tau) = \left(\frac{b}{b_l(\tau)}\right)\left(\frac{l_2(0)}{l_{l(2)}(\tau)}\right), \tag{43b}$$

after calculations of all the Gaussian integrals in Eq. (42), where $l_{l(2)}(\tau) = [2a^2 + b_l^2(\tau)]^{1/2}$ and $l_{coh,(2)}(\tau) = b_l(\tau)[2 + b_l^2(\tau)/a^2]^{1/2}$ are the total and spatial correlation lengths of the two polariton field,

$Z_1 = z_1 + \delta z_l(\tau) - v_l t_o - \int_{t_o}^{t_o+\tau} v(t')dt'$, $Z_2 = Z_1 + z_2 - z_1$, $b_l^2(\tau) = b^2(\tau) + \delta l_{l,l}^2(\tau)$, $b^2(\tau) = b^2 + \int_{t_o}^{t_o+\tau} \omega_{kk}(t')dt'$ (see Appendix 3).

Behavior of the two-polariton function $\Psi_{ll}(t',z';t,z;\tau)$ for different time $\tau$ is shown on the Fig. 2. As seen from Fig. 2, function $\Psi_{ll}(t',z';t,z;\tau)$ describes sharp spatial correlations of two polaritons for $\tau = 0$, which are spreading and disappearing in time at further evolution. Using Eq. (43) we can estimate the condition where the quantum correlations are still preserved. We find the probability of the two polariton detection

$$\tfrac{1}{2}\int_0^L\int_0^L dz'dz|\Psi_{ll}(t,z';t,z;\tau)|^2 = N_2(\tau)/2, \tag{44}$$

which determines the probability of two photon emission from the medium. Similarly to the calculation of Eq. (43 a) we find the second order correlator

$$I_{ll}(t_2,z_2;t_1,z_1;\tau) = \langle \Phi_{l,cor}(t_o;2)|\hat{\Psi}_l^+(t_2,z_2;\tau)\hat{\Psi}_l(t_1,z_1;\tau)|\Phi_{l,cor}(t_o;2)\rangle$$

$$= \frac{4}{\sqrt{\pi}}\frac{l_{l(2)}(0)b}{l_{coh,(1)}(\tau)a^2}\exp\{-\frac{(Z_1^2+Z_2^2)}{2l_{l(1)}^2(\tau)}\}\exp\{-\frac{(Z_1-Z_2)^2}{2l_{coh,(1)}^2(\tau)}\}, \tag{45}$$

where $l_{coh,(1)}(\tau) = \sqrt{2}(l_{l(1)}(\tau)/a^2)\sqrt{l_{l(1)}^2(\tau)l_{l(1)}^2(0) - a^4}]$ is a first order spatial coherence length, $l_{1(l)}(\tau) = [a^2 + b_l^2(\tau)]^{1/2}$ is a longitudinal size of the coherence. We see that the behavior of $\Psi_{ll}(t',z';t,z;\tau)$ and $I_{ll}(t_2,z_2;t_1,z_1;\tau)$ remains similar to each other (see [36]) even time $\tau$ considerably increases. Using Eq. (45) we find an average number of polaritons $N_{pol}$

$$N(\tau) = \int_0^L dz I_{ll}(t,z;t,z;\tau) = \frac{2l_{l(2)}^2(0)b}{\sqrt{l_{l(1)}^2(\tau)l_{l(1)}^2(0) - a^4}}\bigg|_{a \gg b} \cong 2\frac{\sqrt{2}}{(1+b_l^2(\tau)/b^2)^{1/2}} = 2\sqrt{2}N_2(\tau). \quad (46)$$

Temporal behavior of $N_2(\tau)$ and $N(\tau)$ given on the Fig.3 shows a more rapid decay of $N_2(\tau)$. For small spatial spreading $b_l^2(\tau) \ll a^2$ we get the following relation for the coherence lengths: $l_{coh,(1)}(\tau) \cong l_{coh,(2)}(\tau)\sqrt{1+(b/b_l(\tau))^2}$ and $l_{coh,(2)}(\tau) \cong \sqrt{2}b_l(\tau)$, where $l_{coh,(1)}(0) \cong \sqrt{2}l_{coh,(2)}(0)$ for $\tau = 0$ and $l_{coh,(1)}(\tau) \cong l_{coh,(2)}(\tau)$ for a larger spatial spreading $(b/b_l(\tau))^2 \ll 1$. Thus the two polariton field and all the quantum spatial correlations are completely preserved if the spatial spreading satisfies the following important relation:

$$\int_{t_o}^{t} dt' \delta\omega_{kk}^{''}(t') \leq b^2 = (v_l/\gamma_{ph})^2. \quad (47)$$

However the quantum correlations still remain even for larger spatial spreading than takes place in Eq. (47). We find important information about the quantum properties of two-polariton field evolution from the second-order $g_{ll}^{(2)}$ coherence:

$$g_{l,l}^{(2)}(\tau) = \langle g_{l,l}^{(2)}(t,z';t,z;\tau)\rangle = \frac{\langle|\Psi_{l,l}(t,z';t,z;\tau)|^2\rangle}{\langle I_{ll}(t',z';t',z';\tau)\rangle\langle I_{ll}(t,z;t,z;\tau)\rangle} = \frac{b}{2b(\tau)}\frac{\{a^2 + l_{l(1)}^2(0)b_l^2(\tau)/b^2\}}{l_{l(2)}(0)l_{l(2)}(\tau)}, \quad (48)$$

where $\langle...\rangle$ is an averaging over z and z', the $g_{ll}^{(2)}(\tau)$ is shown on the Fig. 4 for $a = 10b$.

As we see from Eq. (48), $g_{ll}^{(2)}$ coherence of the two-polariton field is increased from $g_{ll}^{(2)}(0) = 0.5$ to the classical value $g_{ll}^{(2)}(\tau') = 1$ at $\tau' \cong (3.75b)^2/\delta\omega_{kk}$ (see Fig.4) and then $g_{ll}^{(2)}$ increases asymptotically to the limit $g_{ll}^{(2)}(\tau \to \infty, a = 10b) = a/(2b\sqrt{2}) \cong 3.536$ corresponding again to the nonclassical field (with average number of photons in the state $N(\tau) < N(\tau') = 2b\sqrt{2}/\tilde{b}(\tau') \cong 1.30$ at $\tau > \tau'$). Thus we have found that the original quantum correlations are preserved for time $\tau < \tau'$ that points out a more robust condition for the quantum manipulations of MC-light in comparison with Eq. (47).

Summarizing this part we note that the presented analysis shows the possibility of the transformation of original single color probe quantum pulse to the MC-fields and manipulations of the new field amplitudes preserving the original quantum correlations. We also add that the evaluation of the entanglement in the output MC-field represents an independent interesting subject for the further work, which can be easily performed using the obtained Eqs. (18)-(20) and taking into account the simplified analysis of the fluctuation forces in the spirit of the work [33].

## V. CONCLUSION

The EIT technique for generation and manipulation of the MC-light field from a traveling quantum probe light is proposed. Theoretical analysis of the MC-field dynamics has been performed in the adiabatic limit of interaction between the light fields in the multi double $\Lambda$ coherent atomic medium. In Part II, we have obtained the exact analytical solution of the system of equations for the MC-field evolution in Eqs. (13a) and (13b) where the main formulas are given in Eqs. (18) - (20). The solution demonstrates the possibility of effective MC-light field control in the medium when all the light fields can move with united group velocity which can be effectively controlled and even stopped in the medium by adjusting the laser field amplitudes. In Part III, the spectral conditions for perfect manipulations of the traveling and stationary MC-light fields have been found in terms of minimization of second order dispersion. We have found that the MC-light field evolution follows near dark MC-field state in Eq. (2a). The difference between pure dark state and state of the evolved MC-field is determined by the spatial shifts of different light field envelopes. This difference results in dissipation and spatial spreading of the MC-light fields moving its state nearer to the dark MC-polariton states.

The detail temporal and spatial properties of the MC-field quantum dynamics have been analyzed in part IV for the cases of coherent and two-photon (EPR) correlated state of probe light field. Our results show the possibilities of considerable lifetime lengthening and enhancement of the electric field amplitudes in the stationary MC-light field, which look promising in the improvement of quantum nondemolition measurements and enhancement of weak nonlinearity in the photon-photon and photon-atom interactions with a localized atomic system in spatially limited media. Specific quantum correlations in temporal and spatial dynamics of the two-polariton MC-field have been studied in part IV.c.. It has been shown that the irreversible evolution of the stationary MC-fields is accompanied by the spreading of spatial correlations between the two polaritons given by Eqs. (43) and (45). The conditions have been found where the initial quantum correlations of probe light pulse can be preserved (see Eq.(47) and subsequent analysis).

Finally, we note that the proposed technique of quantum manipulations of the traveling and stationary MC-light fields offers a general approach for unitary control of the complicated multi-frequency light fields. The approach and obtained results, from our research, can be used for analysis of a number of new optical schemes especially in the generation and manipulation of MC-light fields. We also believe that the quantum control of MC-light fields can be potentially interesting for applications in complicated multi-frequency optical systems, where multiple light fields have been involved in controllable switching, and in multi-frequency exchange through perfect quantum interactions.

**ACKNOWLEDGMENTS**

The work was supported by CRI project of MoST of Korea Republic.

### APPENDIX 1:

Taking into account Eqs. (16a), (16b) we obtain

$$f_{m(n),l}(z-z') = (1/2\pi)\int_{-\infty}^{\infty} dk\, e^{ik(z-z')}\{1 + i\eta_{m(n),l}(k;k_o)\}. \tag{1.1}$$

After integration over the k we find

$$f_{m,l}(z-z') = \xi_m\{1 - ik_o\xi_l^{-1}\}\{1 - \tfrac{1}{\xi_l(1-ik_o\xi_l^{-1})}\tfrac{\partial}{\partial z'}\}\{\eta_\chi(z-z')\exp\{-(z-z')(\xi_m - ik_o)\}\}, \tag{1.2}$$

$$f_{n,l}(z-z') = \xi_n\{1 - ik_o\xi_l^{-1}\}\{1 - \tfrac{1}{\xi_l(1-ik_o\xi_l^{-1})}\tfrac{\partial}{\partial z'}\}\{\eta_\chi(z'-z)\exp\{-(z'-z)(\xi_n - ik_o)\}\}, \tag{1.3}$$

where $\eta_\chi(z) = 1, \text{if } (z>0); = 0 \text{ if } z<0$. Assuming that the MC-light fields are localized in the medium so that $\hat{\Psi}_l(t, z=0) = \hat{\Psi}_l(t, L) = 0$ at the medium border we Apply Eqs (1.2) and (1.3) in Eq. (17a) we get

$$\hat{\Psi}_m(t,z') = G_{ml}(z-z') \otimes \hat{\Psi}_l(t,z') = \xi_m\{1-ik_o\xi_l^{-1}\}\int_0^z dz' \exp\{-(z-z')(\xi_m - ik_o)\}(1 + \tfrac{1}{\xi_l(1-ik_o\xi_l^{-1})}\tfrac{\partial}{\partial z'})\hat{\Psi}_l(t,z'). \tag{1.4}$$

$$\hat{\Psi}_n(t,z) = G_{nl}(z-z') \otimes \hat{\Psi}_l(t,z') = \xi_n\{1-ik_o\xi_l^{-1}\}\int_z^L dz' \exp\{-(z'-z)(\xi_n - ik_o)\}\{1 + \tfrac{1}{\xi_l(1-ik_o\xi_l^{-1})}\tfrac{\partial}{\partial z'}\}\hat{\Psi}_l(t,z'), \tag{1.5}$$

Taking into account $\xi_l l_o \gg 1$ in Eqs.(1.4) and (1.5) we simplify the equations to the following

$$\hat{\Psi}_m(t,z)\big|_{\xi_l l_o \gg 1} = G_{ml}(z-z') \otimes \hat{\Psi}_l(t,z') \cong \xi_m\{1-ik_o\xi_l^{-1}\}\int_0^z dz' \exp\{-(z-z')(\xi_m - ik_o)\}\hat{\Psi}_l(t, z' + \tfrac{1}{\xi_l(1-ik_o\xi_l^{-1})}). \tag{1.6}$$

$$\hat{\Psi}_n(t,z)\big|_{\xi_l l_o \gg 1} = G_{nl}(z-z') \otimes \hat{\Psi}_l(t,z') \cong \xi_n\{1-ik_o\xi_l^{-1}\}\int_z^L dz' \exp\{-(z'-z)(\xi_n - ik_o)\}\hat{\Psi}_l(t, z' + \tfrac{1}{\xi_l(1-ik_o\xi_l^{-1})}). \tag{1.7}$$

### APPENDIX 2

Applying Eqs. (16a) and (16b) to Eqs. (13a) and (13b) after all the algebraic calculations we find the resulting fluctuation noise term $\hat{f}_{1,l}^{(4)}(t',k)$ in Eqs. (19a) and (19c) which includes the fluctuation forces effecting field $\hat{\Psi}_l(t,z)$ on all the atomic transitions:

$$\hat{f}_{1,l}^{(4)}(t,k) = \{(1 - \tfrac{\partial}{\mu\partial t})[\tfrac{1}{\mu}\hat{f}_{l;\Sigma}^{(3)}(t,k)] - \hat{f}_{1,l}^{(3)}/\mu\}, \tag{2.1}$$

$$\hat{f}_{l;\Sigma}^{(3)}(t,k) = \mu^{-1}(t)\{\sum_{m=1}^{M_+}\tfrac{\Gamma_m(t)}{(1+i\xi_m^{-1}(k-k_o))}\{\hat{f}_{1,l}^{(3)}(t,k) - \hat{f}_{1,m}^{(3)}(t,k)\} + \sum_{n=1}^{M_-}\tfrac{\Gamma_n(t)}{(1-i\xi_n^{-1}(k+k_o))}\{\hat{f}_{1,l}^{(3)}(t,k) - \hat{f}_{1,n}^{(3)}(t,k)\}, \tag{2.2}$$

$$\hat{f}_{1,m}^{(3)}(t,k) = \{\hat{f}_\Sigma^{(3)}(t,k) - i\tfrac{\mu(t)}{\Omega_m(t)}\hat{f}_{1m}^{(2)}(t,k-k_o)\}, \tag{2.3}$$

$$\hat{f}_{1,n}^{(3)}(t,z) = \{\hat{f}_\Sigma^{(3)}(t,k) - i\tfrac{\mu(t)}{\Omega_n(t)}\hat{f}_{1n}^{(2)}(t,k+k_o)\}, \tag{2.4}$$

$$\hat{f}_\Sigma^{(3)}(t,z) = \mu(t)\int_{t_o}^{t} dt' \exp\{-\int_{t'}^{t} dt''\mu(t'')\}\hat{f}_\Sigma^{(2)}(t',k), \tag{2.5}$$

$$\hat{f}_\Sigma^{(2)}(t,k) = i\{\sum_{m=1}^{M_+}\tfrac{\Omega_{+m}^*(t)}{\tilde{\gamma}_{1,m}}\hat{f}_{1m}^{(2)}(t,k-k_o) + \sum_{n=1}^{M_-}\tfrac{\Omega_{-n}^*(t)}{\tilde{\gamma}_{1,n}}\hat{f}_{1n}^{(2)}(t,k+k_o)\} + \hat{f}_{12}^{(o)}(t,k_o), \tag{2.6}$$

**APPENDIX 3:**

Using the formulas for the group velocity $v(t)$ (21b) and second order dispersion (see Eq. (26) and Eq. (29)) in the optically dense media where $k\xi_{m,n}^{-1} \ll 1$ we can rewrite the functions $J_{m(n)}(t;k,k_o=0)$ in the convenient exponential form taking into account only a second order of smallness of the parameters $k\xi_{m,n}^{-1}$:

$$J_m(t;k,k_o=0) = \frac{1+i\eta_{m;l}(k,k_o)}{\{1+i\beta(t;k,k_o)/\mu(t)\}} \cong 1+i\delta z_m k + k^2\{\tfrac{1}{2}B_l(t)+\xi_m^{-1}(\xi_l^{-1}-\xi_m^{-1})-\delta z_l \delta z_m\} \cong \exp\{i\theta_{m;l}(t,t_o;k)\}, \qquad (3.1)$$

$$J_n(t;k,k_o=0) = \frac{1+i\eta_{n;l}(k,k_o)}{\{1+i\beta(t;k,k_o)/\mu(t)\}} \cong 1+i\delta z_n k + k^2\{\tfrac{1}{2}B_l(t)-\xi_n^{-1}(\xi_l^{-1}+\xi_n^{-1})-\delta z_l(t)\delta z_n(t)\} \cong \exp\{i\theta_{n;l}(t,t_o;k)\}, \qquad (3.2)$$

where $\theta_{m(n);l}(t,t_o;k) = \delta z_{m(n)}(t)k - i\tfrac{1}{2}k^2 \delta l_{m(n),l}^2(t)\}$, $\delta l_{m,l}^2(t) = [\delta z_m(t)(\delta z_m(t)-2\delta z_l(t))+2\xi_m^{-1}(\xi_l^{-1}-\xi_m^{-1})+B_l(t)]$,

$\delta l_{n,l}^2(t) = [\delta z_n(t)(\delta z_n(t)-2\delta z_l(t))-2\xi_n^{-1}(\xi_l^{-1}+\xi_n^{-1})+B_l(t)]\}$, $B_l(t) = 2\frac{1}{\mu(t)}\{\sum_m^{M_+} v_m(t)/\xi_m + \sum_n^{M_-} v_n(t)/\xi_n - v(t)\xi_l^{-1}\}$,

$\delta z_m(t) = [v(t)/\mu(t)-\xi_m^{-1}]$, $\delta z_n(t) = [v(t)/\mu(t)+\xi_n^{-1}]$.

We note that $\delta z_{m,(n)}(t)$ and $\delta l_{m,(n)}^2(t)$ give a kinematical spatial shift and a compression of the field envelopes for the MC-light and these parameters $\delta l_{m,l}^2(t>t_1)=\delta l_{m,l}^2(t>t_1)=0$ and $\delta z_m(t>t_1)=\delta z_n(t>t_1)=0$ after switching to a single frequency traveling slow light field.

# FIGURE'S CAPTIONS

Fig.1. Scheme of the atomic transitions and frequencies of the quantum weak fields $\hat{E}_m$, $\hat{E}_n$ and control laser fields $\Omega_m$, $\Omega_n$ where $m \in [1,...,M_+]$, $n \in [1,...M_-]$.

Figure 2. The two-polariton function $\Psi_{ll}(z';z;\tau)$ given by Eq. (42) at different manipulation time $\tau$ of the stationary MC-light for $a = 10b$, b=1: a) $(\delta\omega_{kk}\tau)^{1/2}/b = 5$, b) $(\delta\omega_{kk}\tau)^{1/2}/b = 20$, c) $(\delta\omega_{kk}\tau)^{1/2}/b = 40$

Fig. 3. Decay of the two-polariton field component ($N_2(\tau)$ -----, Eq. (44)) and average number of polaritons ($N(\tau)$ ———, Eq. (46)) on manipulation time $\tau$ at $a = 10b$, $b = 1$.

Fig. 4. Second-order $g_{ll}^{(2)}(\tau)$ coherence at $a = 10b$, $b = 1$ (see Eq. (48)): $g_{ll}^{(2)}(\tau) < 1$ for $0 < \tau < \tau'$ that corresponds nearly to the initial quantum properties of the MC-field, $g_{ll}^{(2)}(\tau') = 1$ at $\tau' \cong (3.75b)^2/\delta\omega_{kk}$ gives a classical relation for the MC-field; large manipulation time $\tau > \tau'$ leads to the new quantum state of MC-field with $1 < g_{ll}^{(2)}(\tau) < 3.536$.

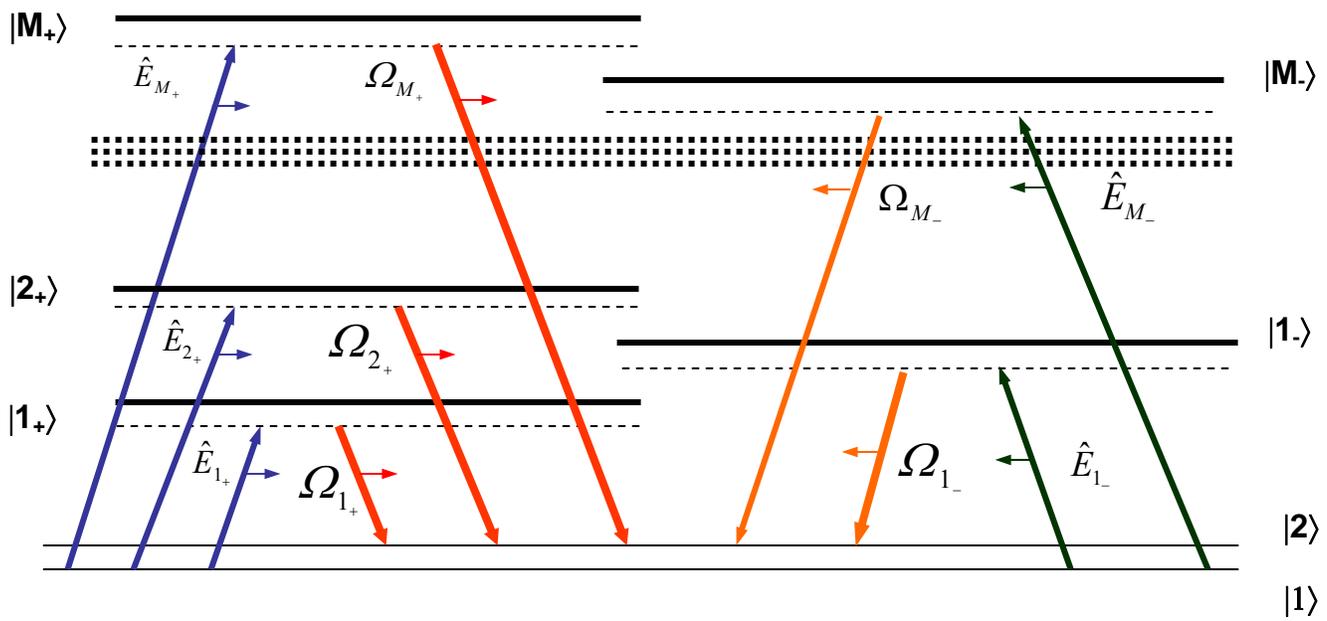

Fig.1.

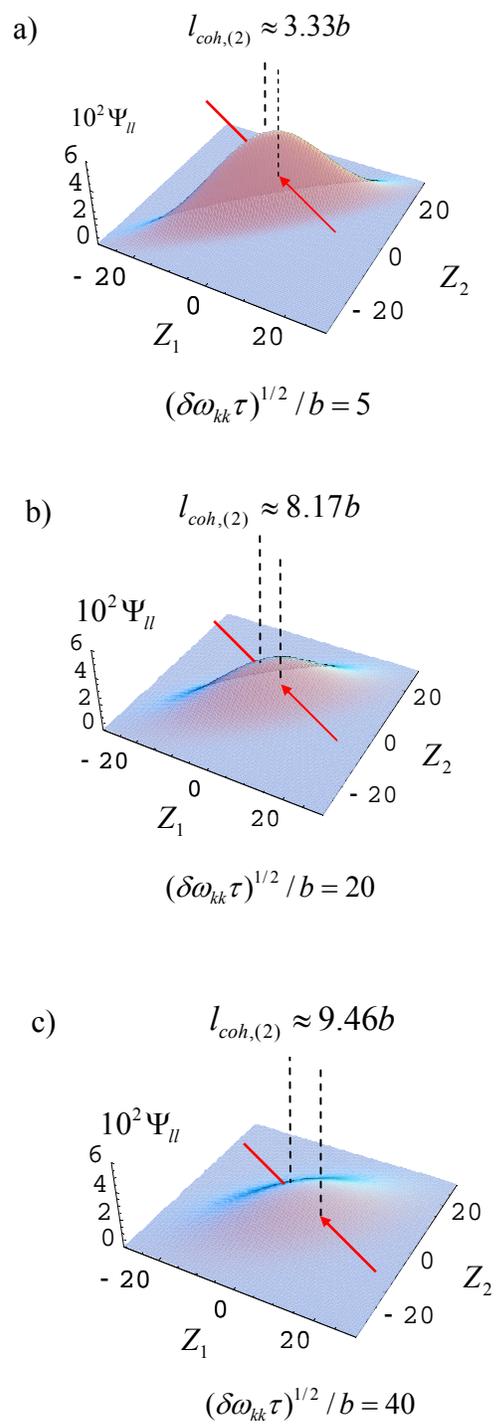

Figure 2.

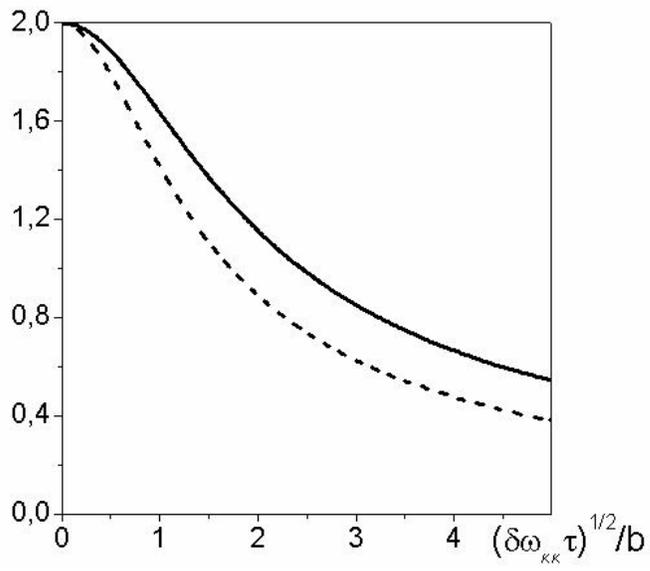

Figure 3.

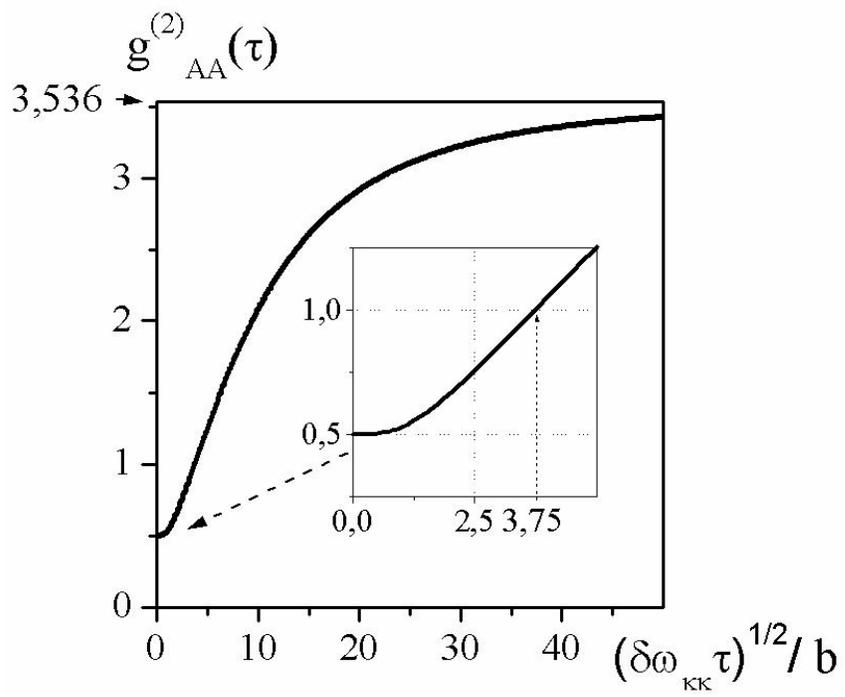

Figure 4.